\documentclass[preprint,prd,aps,showpacs,preprintnumbers,amsmath,amssymb]{revtex4-1}
\usepackage{graphics,epsfig,subfigure}
\usepackage{diagbox}
\usepackage[usenames]{color}
\usepackage{color}
\usepackage{graphicx}
\usepackage{amsfonts}
\usepackage{indentfirst}
\usepackage{booktabs}
\usepackage{hyperref}
\hypersetup{hypertex=ture,backref=true,colorlinks=ture,linkcolor=blue,anchorcolor=blue,citecolor=blue}
\usepackage{float}
\usepackage{multirow}
\usepackage[section]{placeins}
\usepackage{array}
\usepackage{amsmath}
\usepackage{slashed}
\usepackage{caption}
\usepackage{fancyhdr}

\begin{document}
\title{Spherically symmetric horizonless solutions and their frozen states in Bardeen spacetime with Proca field}
\author{Rong Zhang}
\author{Yong-Qiang Wang\footnote{yqwang@lzu.edu.cn, corresponding author}}

\affiliation{ $^{1}$Lanzhou Center for Theoretical Physics, Key Laboratory of Theoretical Physics of Gansu Province,School of Physical Science and Technology, Lanzhou University, Lanzhou 730000, China\\
$^{2}$Institute of Theoretical Physics $\&$ Research Center of Gravitation, Lanzhou University, Lanzhou 730000, China}

\begin{abstract}
In this paper, we construct a static spherical symmetric Bardeen-Proca star (BPS) model, which consists of the electromagnetic field and Proca field minimally coupled with gravity. The introduction of the Proca field disrupts the formation of event horizons, ensuring that these solutions are globally regular throughout the spacetime. We obtain families of BPS solutions under several magnetic charge conditions. Based on these results, we further investigate the ADM mass, Noether charge, and energy density distribution of them. We find that when the magnetic charge is sufficiently large, solutions with a critical horizon $r_{cH}$ emerge as $\omega \rightarrow 0$, and the time component of the metric approaches zero inside $r_{cH}$. To an observer at infinity, the collapse process of the matter near the critical horizon appears frozen. Consequently, we refer to the solution with $r_{cH}$ as the frozen Bardeen-Proca star (FBPS). Additionally, we also investigate the circular geodesic orbits of BPS. For the light ring, we find that the light rings always appear in pairs, located on both sides of the critical horizon and moving further apart as the frequency $\omega$ decreases. For timelike circular orbits, we investigate their distribution in the spacetime of BPSs and highlight four representative families of BPS solutions. 
\end{abstract}

\maketitle
\thispagestyle{empty}

\section{INTRODUCTION}\label{Sec1}
With the advent of a large amount of observational data \cite{EventHorizonTelescope:2019dse, EventHorizonTelescope:2019uob, EventHorizonTelescope:2019jan, EventHorizonTelescope:2019ths, EventHorizonTelescope:2019pgp, EventHorizonTelescope:2019ggy}, people's belief in the existence of black holes has strengthened. However, in General Relativity \cite{Einstein:1916vd}, black holes are always associated with singularities \cite{Schwarzschild:1916uq, Kerr:1963ud, Newman:1965my}, where the curvature of spacetime becomes infinite, causing physical laws to break down. And the singularity theorems proved by Penrose and Hawking \cite{Penrose:1964wq, Hawking:1970zqf} indicate that if the strong energy condition and global hyperbolic spacetime assumption are satisfied, singularities are unavoidable. However, singularities cannot exist in the real physical world and are merely a manifestation of the limitations of general relativity.

To avoid singularities, various models have been proposed to regularize black holes. The study of regularizing black holes can be traced back to the work of Shirokov \cite{Shirokov1948} and Duan \cite{Duan:1954bms}, who attempted to regularize the Reissner-Nordström metric. Unfortunately, these results were largely overlooked for decades. Then in the 1960s, Sakharov \cite{Sakharov:1966aja} and Gliner \cite{Gliner:1966} proposed that a de Sitter metric at the center could avoid the singularity. Building on their work, Bardeen \cite{Bardeen:1968} proposed the first regular black hole model by replacing the Schwarzschild black hole mass with a $r$-dependent function, which is referred to the Bardeen black hole. Over thirty years later, Beato and Garcia discovered the matter source of the Bardeen black hole model, interpreting it as a magnetic monopole in the context of nonlinear electrodynamics \cite{Ayon-Beato:2000mjt}. In recent years, various regular black hole models \cite{Dymnikova:1992ux,Ayon-Beato:1998hmi,Balart:2014cga,Balart:2016zrd,Rodrigues:2018bdc,deSousaSilva:2018kkt,Balart:2014jia,Bambi:2013ufa,Lan:2020fmn,Bueno:2024dgm,Barenboim:2024dko} have been constructed using different nonlinear electromagnetic fields as sources (see Refs. \cite{Ansoldi:2008jw,Lan:2023cvz,Torres:2022twv,Carballo-Rubio:2025fnc} for reviews). These models circumvent the singularity theorems by introducing matter that violates the strong energy condition.

For matter fields other than nonlinear electromagnetic fields, certain physical mechanisms can also avoid the formation of singularities. Boson stars are one widely studied example \cite{Li:2020ffy,Herdeiro:2020jzx,Sun:2022duv,Zhang:2021xhp,Kleihaus:2009kr}. They are composed of massive bosons, whose self-gravity is balanced by pressure arising from the Heisenberg uncertainty principle. Because their mass can be substantial, they are considered feasible black hole alternatives in certain cases \cite{Guzman:2009zz,Cunha:2015yba,Cardoso:2019rvt}. The concept of boson stars can be traced back to Wheeler's attempt to construct stable solitons (known as ``geons") using classical electromagnetic fields, which was unsuccessful. Subsequently, Kaup \cite{Kaup:1968zz} and Ruffini \cite{Ruffini:1969qy} replaced the electromagnetic fields with the massive scalar fields, successfully constructing stable soliton solutions, which were the first boson star model. Since then, research on boson stars has also expanded to include massive vector fields \cite{Brito:2015pxa,SalazarLandea:2016bys,Su:2023zhh,Zhang:2023rwc} and spinor fields \cite{Finster:1998ws,Finster:1998ux,Dzhunushaliev:2018jhj,Huang:2023glq,Hao:2023igi}, which are referred to as Proca stars and Dirac stars, respectively.

Recent research has found that introducing matter fields into the Bardeen theory limits the value of magnetic charge. Therefore, the new model neither produces singularities nor forms event horizons. For instance, Ref. \cite{Wang:2023tdz} coupled the Bardeen action with a complex scalar field, resulting in the Bardeen-boson stars (BBSs). Essentially, this model is a boson star with magnetic charge. When the magnetic charge is small, the properties of the BBSs are similar to pure boson star. However, when the magnetic charge is sufficiently large, the scalar field frequency can approach zero, concentrating the scalar field almost entirely within a critical radius $r_{cH}$. At $r = r_{cH}$, the value of $1/g_{rr}$ is close to zero, but it does not actually reach zero, therefore, this radius $r_{cH}$ is not a true horizon and is referred to as a critical horizon. Inside the critical horizon, $g_{tt}$ approaches zero, making the collapse process of the matter approaching $r_{cH}$ appears frozen to an observer at infinity. Hence, these solutions are known as frozen Bardeen-boson stars (FBBSs). This model can be further generalized, such as by replacing the Bardeen action with Hayward action \cite{Yue:2023sep, Chen:2024bfj}, using other matter fields \cite{Huang:2023fnt, Huang:2024rbg, Zhang:2024ljd, Sun:2024mke, Zhao:2025hdg}, or considering modified gravity \cite{Ma:2024olw,Wang:2024ehd}.

In these studies, the Proca field was not explored. However, Proca stars exhibit several noteworthy properties, particularly their enhanced stability compared to scalar boson stars\cite{Sanchis-Gual:2017bhw,Sanchis-Gual:2019ljs,DiGiovanni:2020ror}. Furthermore, recent studies based on Proca star models applied to observational data have further strengthened our confidence in their physical existence \cite{CalderonBustillo:2020fyi,Rosa:2022toh,Rosa:2022tfv}. Motivated by these findings, this paper investigates the Bardeen spacetime with the Proca field, which we refer to as Bardeen-Proca stars (BPSs). We find that the BPS exhibits similar properties to the Bardeen-Boson star, including the absence of singularities and event horizons under any conditions, and can produce frozen star solutions under specific circumstances. We also study the ADM mass, Noether charge, and circular geodesic orbits of the BPSs, and analyze the influence of magnetic charge on these properties.

The structure of this paper is as follows. In Sec. \ref{sec2}, we construct Bardeen-Proca stars, consisting of an electromagnetic field and Proca field minimally coupled with gravity. In Sec. \ref{sec3}, we present the boundary conditions for the equations of motion. Sec. \ref{sec4} displays the numerical results of Bardeen-Proca stars and analyzes their properties. Finally, in Sec. \ref{sec5}, we summarize our research findings and discuss potential directions for future work.

\section{THE MODEL}\label{sec2}
\subsection{Framework}
First, we consider the system of Einstein gravity minimally coupled to the Proca field and the nonlinear electromagnetic field of the Bardeen model. The action of the model is given by:
	\begin{equation}
		S=\int \sqrt{-g} d^4 x \left( \frac{R}{16\pi G} + {\cal L}^{B} + {\cal L}^{P}\right),
		\label{eq:action}
	\end{equation}
where $R$ is the Ricci scalar, $G$ is the gravitational constant, and ${\cal L}^{B}$ and ${\cal L}^{P}$ represent the Lagrangian density of the electromagnetic field and the Proca field, respectively. Their explicit forms are:
	\begin{equation}
		\mathcal{L}^{B}=-\frac{3}{2s}(\frac{\sqrt{2q^2 \mathcal{B}}}{1+\sqrt{2q^2\mathcal{B}}})^{5/2},  
		\label{eq:lagrangianb}
	\end{equation}  
	
	\begin{equation}
		\mathcal{L}^{P}= -\frac{1}{4}{A}_{\alpha\beta}\bar{{A}}^{\alpha\beta}
		-\frac{1}{2}\mu^2{A}_\alpha\bar{{A}}^\alpha .
		\label{eq:lagrangiandp}
	\end{equation}  
Here $\mu$ is the mass of the Proca field. The term $\mathcal{B}=\frac{1}{4}{B}_{\alpha \beta} {B}^{\alpha \beta}$ is defined with the electromagnetic field strength ${B}_{\alpha \beta}=\partial_{\alpha}{E}_{\beta}-\partial_{\beta}{E}_{\alpha}$, where ${E}$ is the electromagnetic field. The term $\mathcal{F}=\frac{1}{4}F_{\alpha \beta}F^{\alpha \beta}$ is defined with the Proca field strength $F_{\alpha \beta}=\partial_{\alpha}A_{\beta}-\partial_{\beta}A_{\alpha}$, where $A_{\alpha}$ is the 4-potential. $q$ and $s$ are parameters, where $q$ represents the magnetic charge. The field equations are:
	\begin{equation}
		R_{\alpha \beta} - \frac{1}{2}g_{\alpha \beta}R - 8\pi G (T^{B}_{\alpha \beta}+T^{P}_{\alpha \beta})=0,
		\label{eq:einstein}
	\end{equation}
	\begin{equation}
		\nabla_\alpha \left(\frac{\partial \mathcal{L}^{B}}{\partial \mathcal{B}} {B}^{\alpha \beta}\right)=0,
		\label{eq:equationb}
	\end{equation}
	\begin{equation}
		\nabla_\alpha F^{\alpha \beta} = {\mu}^2 A^\beta .
		\label{eq:equationp}
	\end{equation}
The energy-momentum tensors of the electromagnetic field and the Proca field are respectively:
\begin{equation}
	T_{\alpha \beta}^{B}=-\frac{\partial \mathcal{L}^{B}}{\partial \mathcal{B}} {B}_{\alpha \sigma} {B}_{\beta}^{~\sigma}+g_{\alpha \beta} \mathcal{L}^{B},
	\label{eq:energymomb}
\end{equation}
\begin{equation}
	T_{\alpha \beta}^{P}=\frac{1}{2} ( {F}_{\alpha \sigma }\bar{F}_{\beta \gamma} +\bar{F}_{\alpha \sigma } {F}_{\beta \gamma})g^{\sigma \gamma} -\frac{1}{4}g_{\alpha \beta} {F}_{\sigma\tau}\bar{F}^{\sigma\tau}+\frac{1}{2}\mu^2\left[{A}_{\alpha}\bar{{A}}_{\beta}+\bar{{A}}_{\alpha}{A}_{\beta}-g_{\alpha \beta} {A}_\sigma\bar{{A}}^\sigma\right]  .
	\label{eq:energymomp}
\end{equation}

Considering that the action of the Proca field is invariant under the global $ U(1)$ transformation ${A}_{\alpha}\rightarrow e^{i\zeta}{A}_{\alpha}$, where $\zeta$ is a constant. Therefore, there should be a corresponding conserved current in this system
\begin{equation}
	j^\alpha=
	\frac{i}{2}\left[\bar{F}^{\alpha\beta}{A}_\beta-{F}^{\alpha \beta}\bar{{A}}_\beta\right] .
\end{equation}
Integrating the timelike component of the conserved current over a spacelike hypersurface $\varSigma$ gives the Noether charge $Q$ of the Proca field:
\begin{equation}
	Q=\int_\varSigma \sqrt{-g} j^{\mathrm{t}}dV,
	\label{eq:generalcharge}
\end{equation}
where $ j^{\mathrm{t}}$ is the time component of the current. 

ADM mass is also an important physical quantity. Similar to the Noether charge $Q$, the ADM mass $M$ can be obtained by integrating the Komar energy over a spacelike hypersurface:
\begin{equation}
	M=\int_\varSigma{\sqrt{-g}\left( 2 T_{t}^{t}-T_{\alpha}^{\alpha} \right)dV} .\label{KormarEnergy}
\end{equation}

\subsection{Ansatz}
In this study, we focus on spherically symmetric static configurations, which can be described using the following metric ansatz:
\begin{equation}
	d s^2=-n(r)o^2(r) d t^2+\frac{1}{n(r)} d r^2+r^2\left(d \theta^2+\sin ^2 \theta d \varphi^2\right) .
	\label{eq:metric}
\end{equation}
where $n(r)= 1-\frac{2m(r)}{r}$, $m(r)$ and $o(r)$ only depend on the radial distance $r$. Furthermore, we adopt the following ansatz for the electromagnetic and Proca fields:
\begin{equation}
	{E}=q \cos (\theta) d \varphi,
	\label{eq:anastzb}
\end{equation}
\begin{equation}
	{A}=\left[f(r)dt+ig(r)dr\right] e^{-i \omega t}  ,
	\label{eq:anastzp}
\end{equation}
where $f(r)$ and $g(r)$ are real functions of $r$, and $\omega$ is the frequency of the Proca field.
By substituting the above ansatzs (\ref{eq:metric}--\ref{eq:anastzp}) into Eqs. (\ref{eq:einstein}--\ref{eq:equationp}), we can derive the field equations for $m(r)$, $o(r)$, $f(r)$ and $g(r)$.
These equations can be solved after imposing appropriate boundary conditions.

\subsection{Circular orbits in BPSs spacetimes}\label{c}
The light ring and accretion disk are two important observable features of black holes. In recent years, several studies have investigated the similarities and differences in observable phenomena between Proca stars and black holes. \cite{Herdeiro:2021lwl,Sengo:2024pwk}. These studies have found that under certain conditions, Proca stars appear the light rings, and can mimic disk’s inner edge without the innermost stable circular orbits (ISCO). For Bardeen-Proca stars, we are also interested in their observable features. In order to study the light ring and accretion disk of them, in this subsection, we will consider the basic equations to compute both the light rings (LRs) and timelike circular orbits (TCOs) in the BPSs spacetime.

For a particle moving in the spacetime described by the metric (\ref{eq:metric}), its Lagrangian can be expressed by the following equation:
    \begin{eqnarray}
     2\mathcal{L}=g_{\alpha\beta}\frac{dx^{\alpha}}{d\tau}\frac{dx^{\beta}}{d\tau}=-n(r)o^2(r) \dot{t}^2 +\frac{1}{n(r)} \dot{r}^2 + r^2 \sin  ^2 \theta \dot{\varphi}^2 =\delta. \label{Lequation}
    \end{eqnarray}
Here $\tau$ is the affine parameter of the geodesic, the dot represents the derivative with respect to $\tau$, for null (timelike) geodesics $\delta=0~(\delta=+1)$. The corresponding generalized momentum is
    \begin{eqnarray}
     	p_t=-\frac{\partial \mathcal{L}}{\partial \dot{t}}=-g_{tt} \dot{t},~ 
 		p_{\varphi}= \frac{\partial \mathcal{L}}{\partial \dot{\varphi}}=g_{\varphi \varphi} \dot{\varphi},~
 		p_r=\frac{\partial \mathcal{L}}{\partial \dot{r}}=g_{rr} \dot{t}. \label{Gmotumequation}
    \end{eqnarray}
Substituting Eq. (\ref{Lequation}) into the Lagrange equation yields
 \begin{equation}
 	\frac{dp_t}{d \tau}=-\frac{\partial \mathcal{L}}{\partial t}=0,~ \frac{dp_\varphi}{d \tau}=\frac{\partial \mathcal{L}}{\partial \varphi}=0.
 \end{equation}
Thus $p_t$ and $p_\varphi$ are conserved quantities, representing the energy and the angular momentum, respectively, which can be denoted as $E$ and $L$. 

Since the system we are studying has spherical symmetry, we can simply focus on particles in the equatorial plane $(\theta=\frac{\pi}{2})$. Substituting $E=-g_{tt} \dot{t}$ and $L=g_{\varphi \varphi} \dot{\varphi}$ into Eq. (\ref{Lequation}), we obtain
 \begin{equation}
 	\dot{r}^2 =n(\frac{E^2}{no^2} - \frac{L^2}{r^2}-\delta),
 	\label{eq:Veff}
 \end{equation}
we denote $\dot{r}^2$ with $V(r)$. For particles in a circular orbit with $r=r_{\delta}$, the following condition must be satisfied:
 \begin{equation}
 	V(r_{\delta})=0,~~V'(r_{\delta})=0. \label{effective}
 \end{equation}
Here the symbol $'$ indicates the derivative with respect to the radial coordinate $r$.

Let us first consider null geodesics ($\delta=0$). From Eq. (\ref{eq:Veff}), we obtain 
\begin{equation}
    \dot{r}^2+\frac{L^2}{o^2}(\frac{E^2}{L^2}- \frac{no^2}{r^2})=0.
\end{equation}
Since $E$ and $L$ are conserved quantities and $\dot{r}=0$ for a LR, we can define the effective potential $V_{0}(r)$ for null geodesics as follows:
\begin{equation}
     V_{0}=\frac{no^2}{r^2}.
     \label{eq:VEFF}
\end{equation}
From Eq. (\ref{effective}), we can deduce that the position of a light ring $r_{LR}$ is determined by the effective potential, which should satisfy
\begin{equation}
    \left.\frac{dV_{0}}{dr}\right|_{r_{0}}=0.
    \label{eq:Deffective}
\end{equation}
Additionally, the stability of the light ring is determined by the second derivative of the effective potential. If $V''_{0}(r_{LR})>0$, the light ring is stable. Conversely, if $V''_{0}(r_{LR})<0$, indicating the unstable light ring.

Let us now turn to timelike geodesics ($\delta=+1$). From Eq. (\ref{eq:Veff}), we obtain 
\begin{equation}
   \dot{r}^2o^2 =E^2- no^2(\frac{L^2}{r^2}+1).
   \label{eq:VEFFTL}
\end{equation}
Similar to the analysis of null geodesics, we can define an effective potential $V_{1}$ for timelike geodesics:
\begin{equation}
	V_{1}=no^2(\frac{L^2}{r^2}+1).
    \label{eq:VEFFT}
\end{equation}
The position of a timelike circular orbit $r_{TL}$ is determined by $V_{1}$, which satisfies
\begin{equation}
    \left.\frac{dV_{1}}{dr}\right|_{r_{TL}}=0.
    \label{eq:Deffective}
\end{equation}
Similarly, the stability of TCOs is also determined by the second derivative of the effective potential. If $V''_{1}(r_{TL})>0$, TCOs are stable. Conversely, if $V''_{1}(r_{TL})<0$, indicating the Unstable TCOs.

Moreover, Eqs. (\ref{eq:VEFFT}, \ref{eq:Deffective}) indicate that the position of the TCOs could not be determined by the background spacetime metric solely, but their existence can nonetheless be confirmed. From Eq. (\ref{effective}), we obtain expressions of $E^2$ and $L^2$ for TCOs \cite{Cardoso:2008bp}:
\begin{equation}
	E^2=-\frac{2(no^2)^2}{2no^2-r(no^2)'},~L^2=\frac{r^3no^2}{2no^2-r(no^2)'}.
    \label{eq:EL}
\end{equation}
For TCOs to exist, the energy and angular momentum of particles on orbits must be real. Therefore, in regions that satisfy $(no^2)'<0$ or $2no^2-r(no^2)'<0$, no TCOs exist. Furthermore, from Eq. (\ref{eq:Deffective}), circular null geodesics satisfy: $2no^2-r(no^2)'|_{r_{0}}=0$. Therefore, we can consider that the LRs are located at the edge of the region without TCO. For a more detailed discussion on the relationship between LRs and TCOs, we refer the reader to Ref. \cite{Delgado:2021jxd}.

\section{BOUNDARY CONDITIONS AND NUMERICAL METHODS}\label{sec3}
Before solving the field Eqs. (\ref{eq:metric}--\ref{eq:anastzp}) based on the ansatzs (\ref{eq:metric}--\ref{eq:anastzp}), we need to choose appropriate boundary conditions. Since the Bardeen-Proca stars spacetime is asymptotically flat, the field functions should satisfy the following boundary conditions at infinity:
\begin{equation}
	\qquad m(\infty) = M,\qquad o(\infty) = 1,\qquad f(\infty) = 0, \qquad g(\infty) = 0.
\end{equation}
At the origin, we impose regularity boundary for the metric and Proca field
\begin{equation}
	m(0) = 0,\qquad o(0) = o_0,\qquad \left.\frac{df(r)}{dr}\right|_{r = 0} = 0, \qquad g(0) = 0.
\end{equation}
$M$ and $o_0$ are constants, which can be determined by solving field equations.

The solution of BPSs depends on six input parameters: the gravitational constant $G$, the Proca field mass $\mu$, the magnetic charge $q$ and coupling parameter $s$ in the Bardeen action, where $q$ and $s$ are dimensionless quantities. For convenience of calculation, we can introduce the following dimensionless transformations:
\begin{eqnarray}
        r\rightarrow r\mu,\quad\omega\rightarrow\omega/\mu,\quad  f\rightarrow \frac{\sqrt{4 \pi }}{M_{PL}}f ,\quad g\rightarrow \frac{\sqrt{4 \pi }}{M_{PL}}g.
\end{eqnarray}
where $M_{PL}\equiv 1/\sqrt{G}$ is the Planck mass. Consequently, the dependence on both $G$ and $\mu$ disappears from the equations. In subsequent calculations, we set $s=0.3$. Therefore, the solutions we obtain are determined by only $\omega$ and $q$.

Additionally, to facilitate solving, we also need to introduce the following conformal transformation
    \begin{equation}
        x = \frac{r}{1+r} \label{eq-radius}
    \end{equation}
to map the range of radial coordinates $r\in[0,+\infty )$ into $x\in[0,1]$.

Our solving process is based on the finite element analysis methods, which solves weak form of partial differential equations to obtain numerical results. The iteration process follows the Newton-Raphson method. In general, we set the number of grid points within the integration region $x\in[0,1]$ equals 10000. Using the above numerical method, the relative error of the obtained results is generally below $1\times 10^{-5}$.

\section{NUMERICAL RESULTS}\label{sec4}
In this section, we present results of the BPSs, along with an analysis of their several important quantities. Before presenting our numerical results, it is important to consider two special cases. The first case refers to when $q=0$, where BPSs degenerate into the Einstein-Proca model, that is, the Proca stars. Another case refers to when $q \neq 0$ and the Proca field vanishes, where BPSs degenerate into the Bardeen model, which can describe either a horizonless Bardeen spacetime or a Bardeen black hole, depending on the value of the magnetic charge $q$. The metric of Bardeen model is the following form
\begin{equation}
	ds^2=-k(r)dt^2+k(r)^{-1}dr^2+r^2\left(d\theta^2+\sin^2\theta d\varphi^2\right),~~k(r)=1-\frac{q^3r^2}{s(r^2+q^2)^{3/2}}.
	\label{eq:BardeenElement}
\end{equation}
By solving the equation $k(r) = 0$, it is clear that when $q < 3^{3/4}\sqrt{s/2}$, there is no event horizon in the Bardeen spacetime, therefore it is not a black hole solution. Conversely, when $q \geq 3^{3/4}\sqrt{s/2}$, the Bardeen model has one or two event horizons, so the line element (\ref{eq:BardeenElement}) describes a Bardeen black hole. Especially noteworthy is the case of $q = 3^{3/4}\sqrt{s/2}$, where the equation $k(r) = 0$ has only one root. We refer to the Bardeen black hole with $q=3^{3/4}\sqrt{s/2}$ as the extreme Bardeen black hole. Additionally, $q=3^{3/4}\sqrt{s/2}$ also holds special significance for BPSs. Based on our numerical results of BPSs, we find that the magnetic charge $q$ can only be less than $3^{3/4}\sqrt{s/2}$, otherwise no solution exists. In our paper, we set $s=0.3$. Therefore the magnetic charge of BPSs is restricted to $q<0.883$ in our subsequent discussions.
\begin{figure}[!htbp]
	\begin{center}
    	\includegraphics[height=.28\textheight]{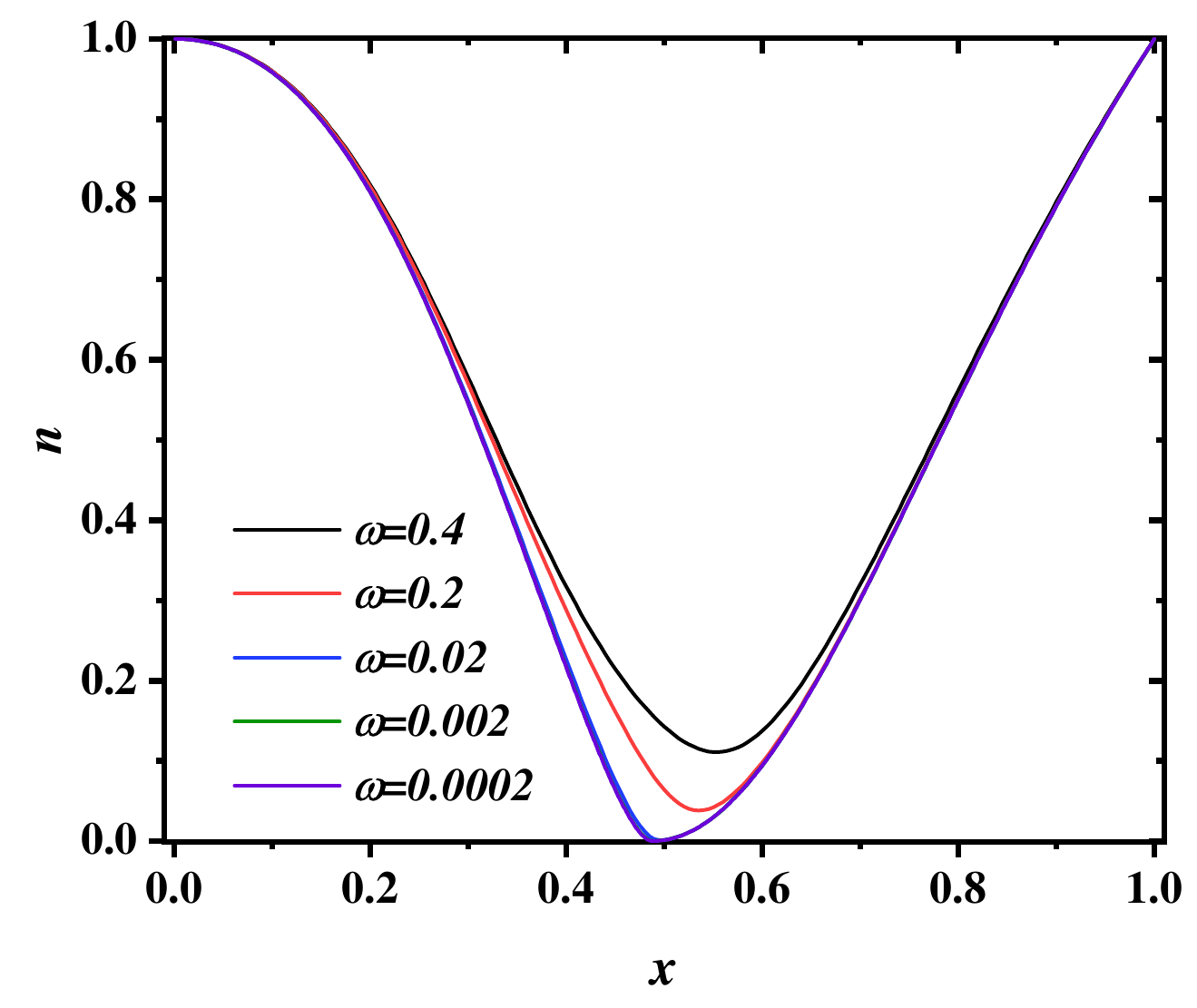}
		\includegraphics[height=.28\textheight]{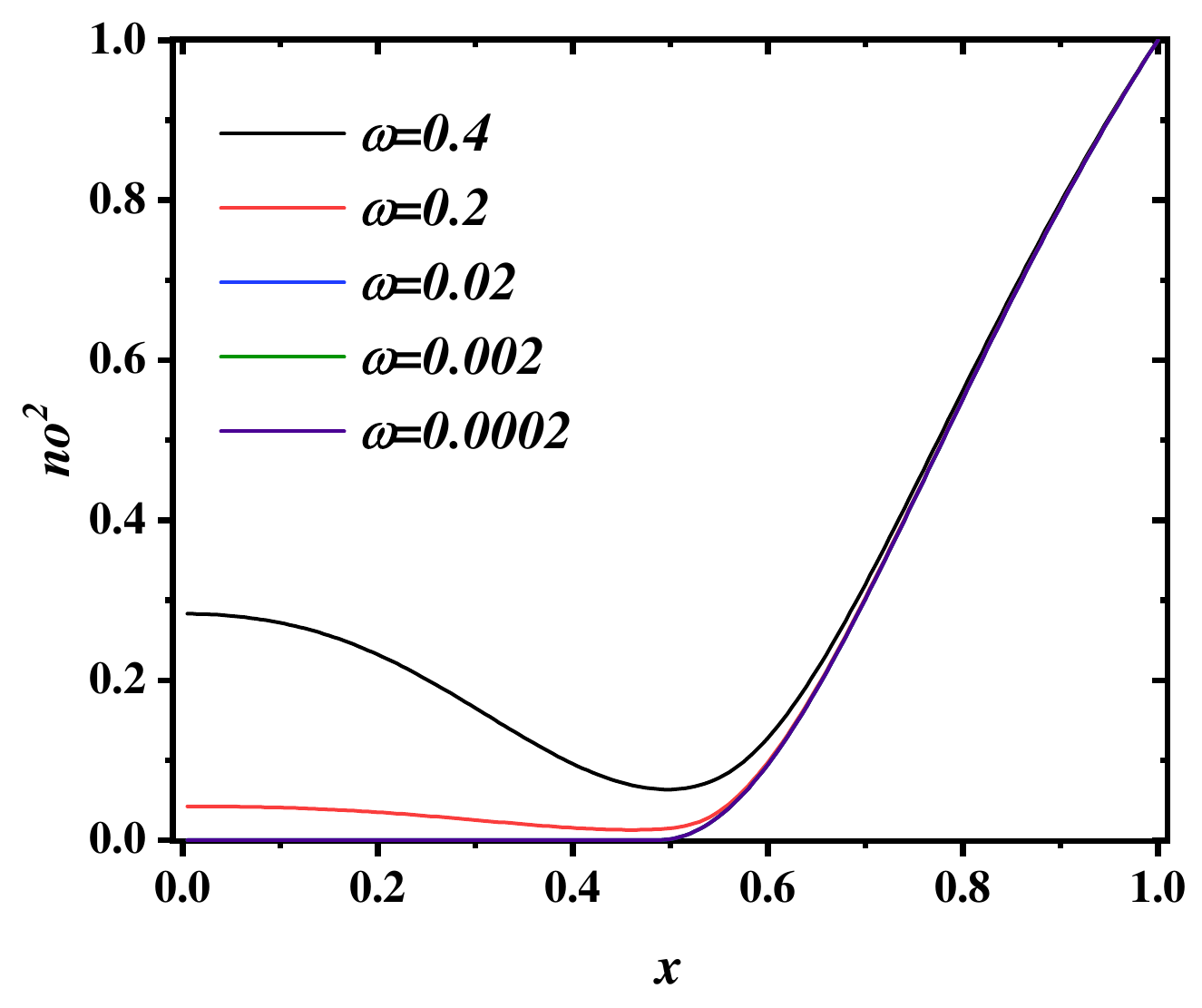}
		\includegraphics[height=.28\textheight]{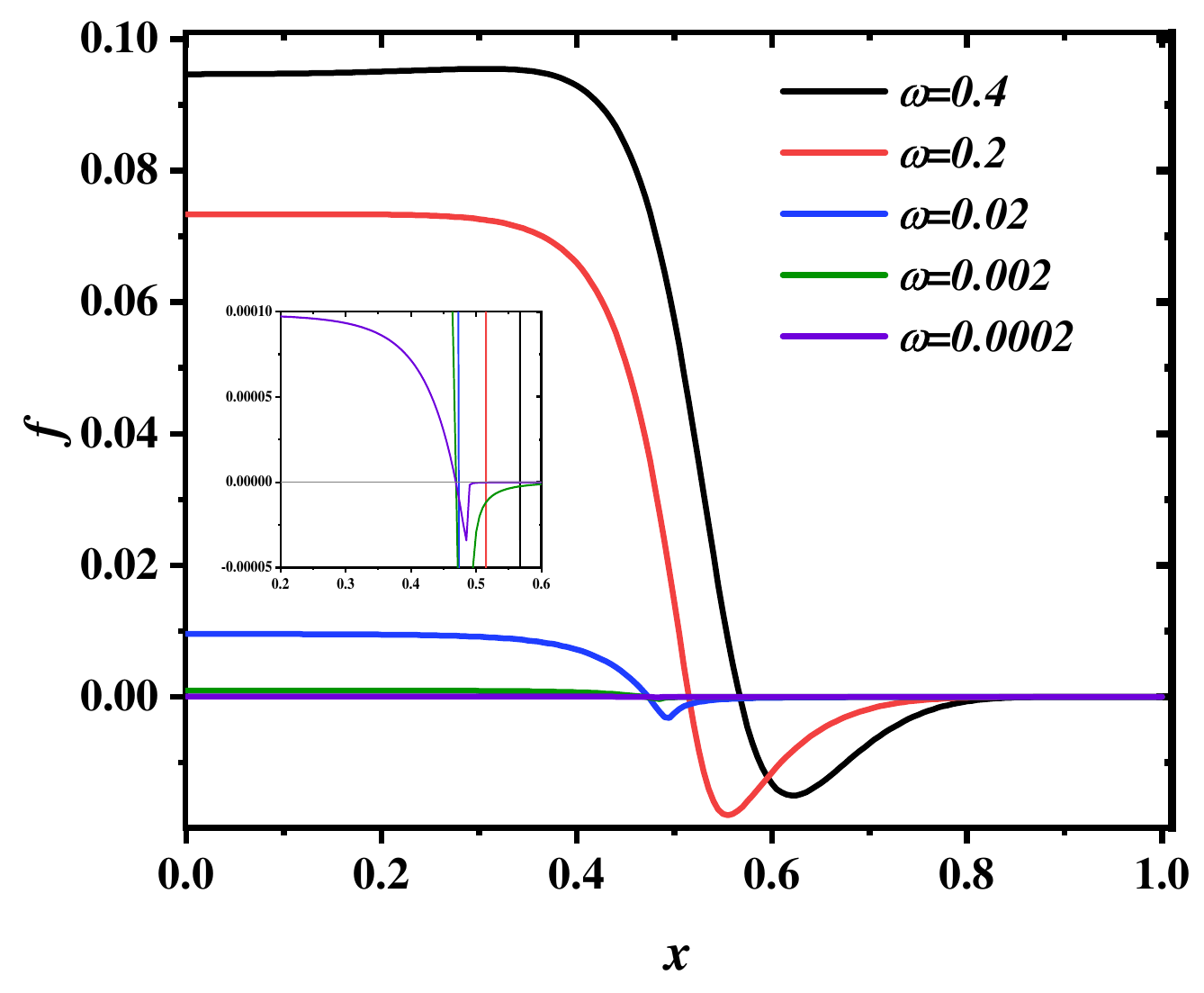}
		\includegraphics[height=.28\textheight]{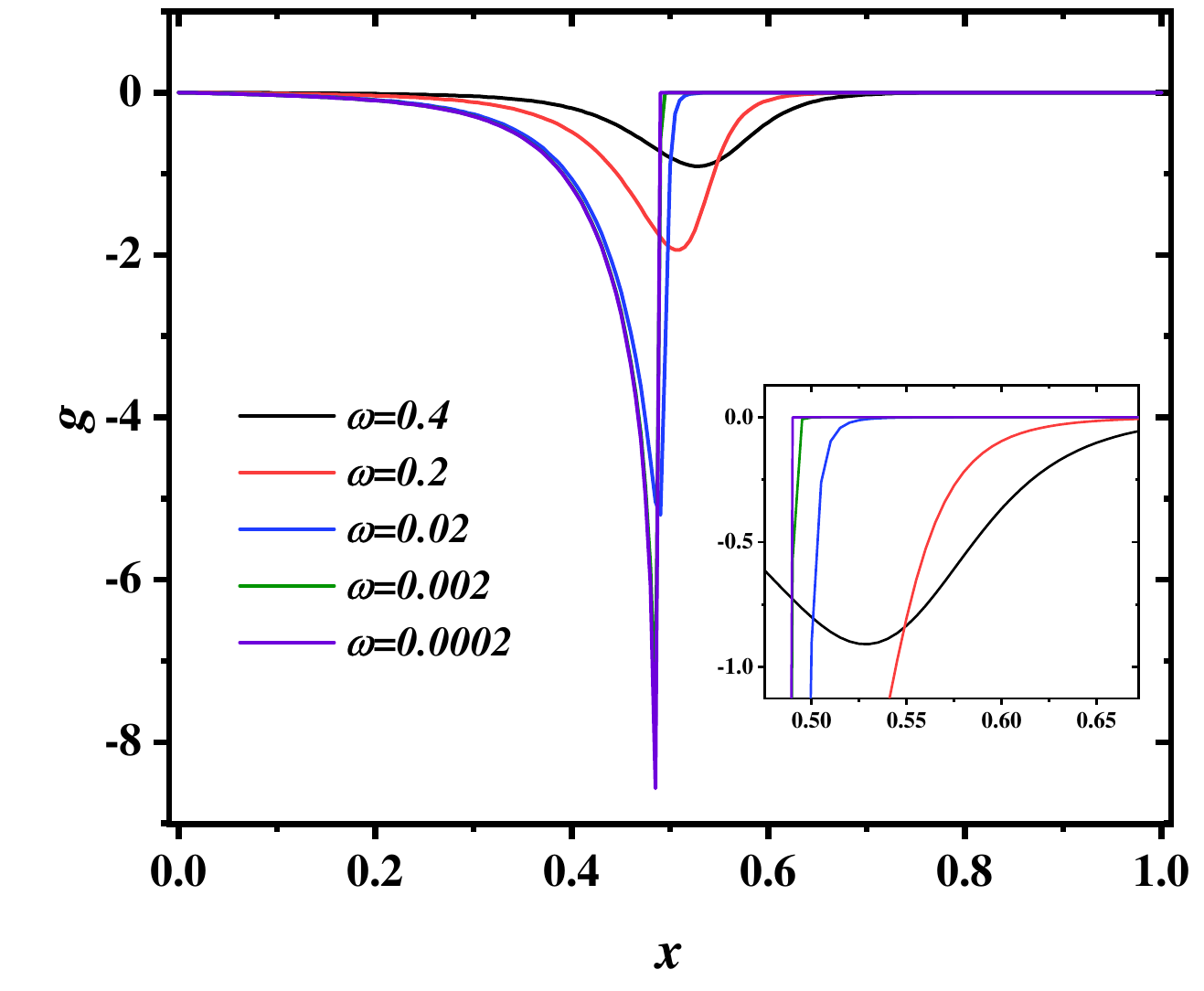}

	\end{center}
	\caption{The Proca metric functions $n$ and $no^2$, field functions $f$ and $g$ of BPSs as functions of $x$, the magnetic charge $q$ of BPSs is fixed at $0.8$.}
	\label{field}
\end{figure}

\begin{figure}[!htbp]
	\begin{center}
		\includegraphics[height=.28\textheight]{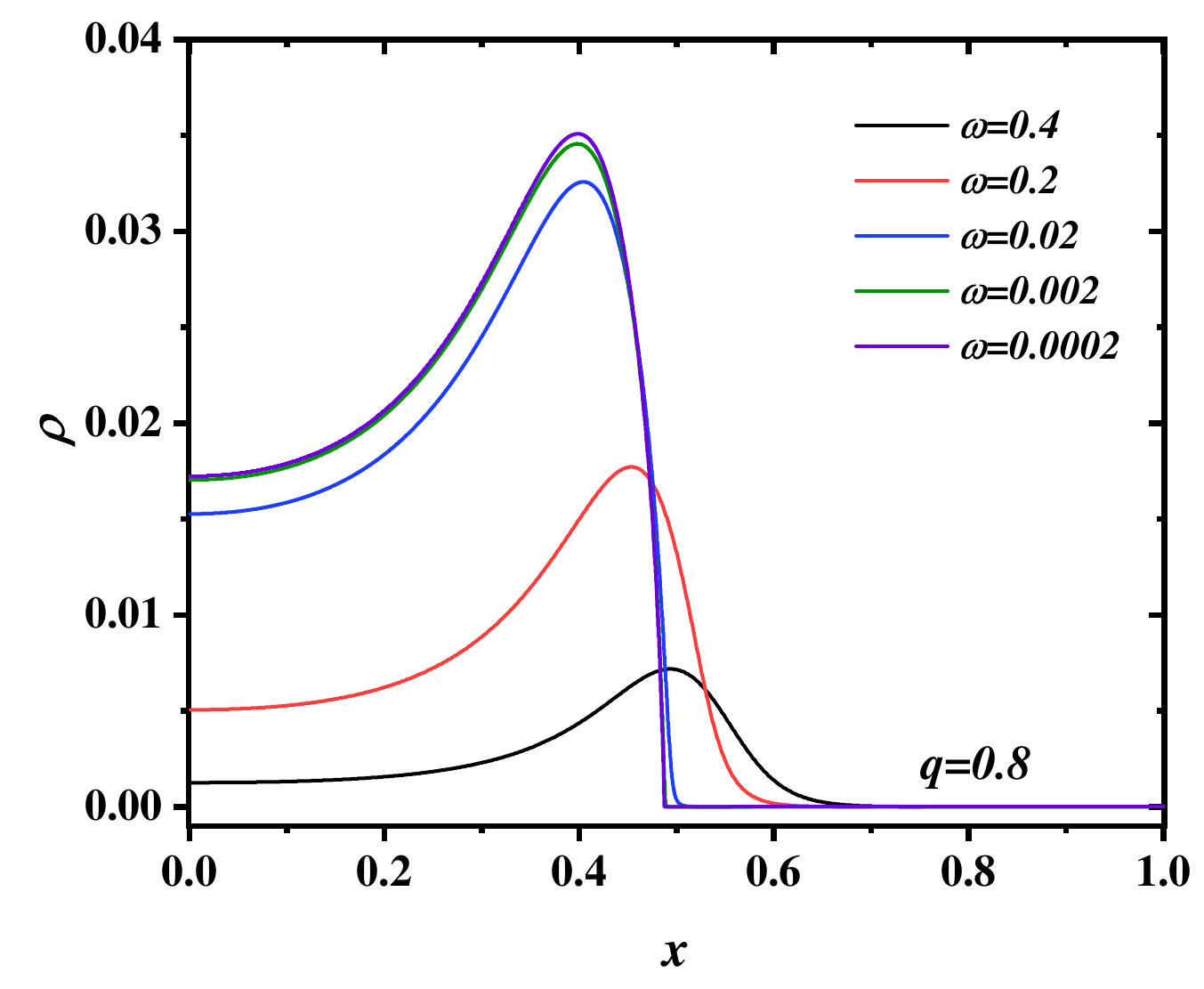}
		\includegraphics[height=.28\textheight]{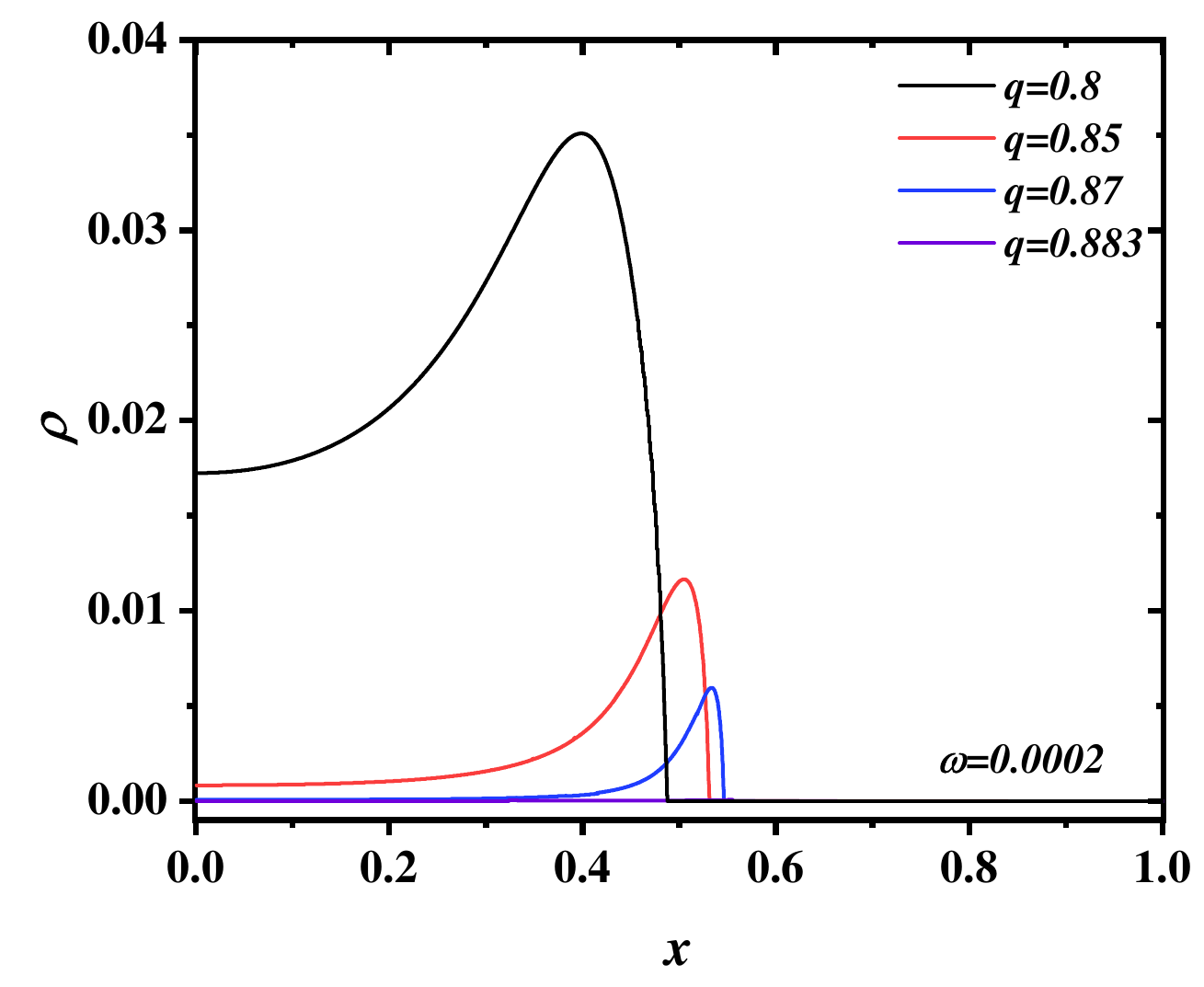}
	\end{center}
	\caption{Left panel: The energy density $\rho$ of BPSs with different values of frequency $\omega$ as functions of $x$, the magnetic charge $q$ is fixed at $0.8$. Right panel: The energy density $\rho$ of FBPSs with different values of magnetic charge $q$ as functions of $x$, the field frequency $\omega$ is fixed at $0.0002$.}
	\label{energy}
\end{figure} 

\begin{figure}[!t]
\begin{center}
\includegraphics[height=.28\textheight]{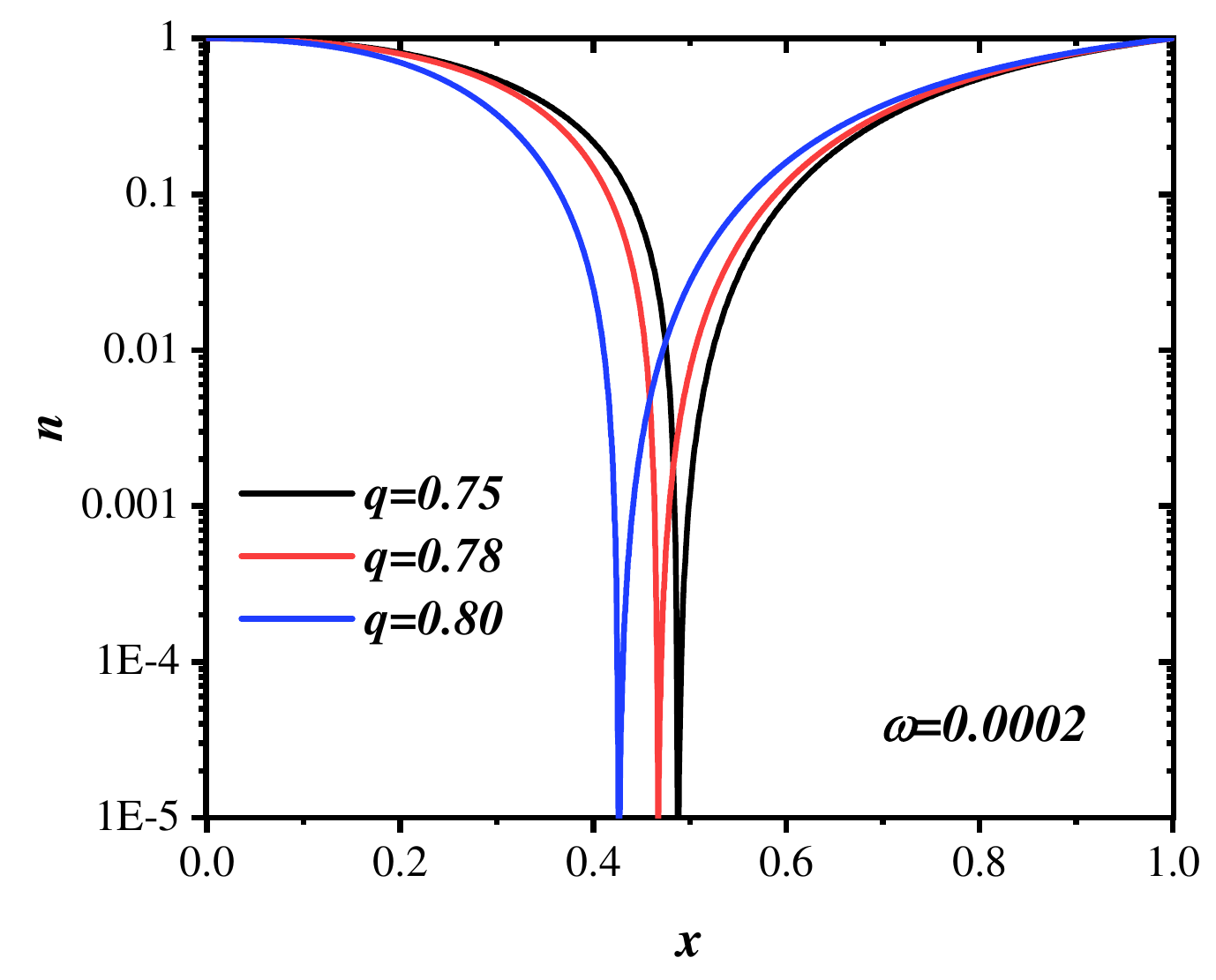}
\end{center}
\caption{The metric functions $n$ of FBPSs as functions of $x$ with different values of magnetic charge $q$, the frequency $\omega$ of BPSs is fixed at $0.0002$.}
\label{lognx}
\end{figure}

As mentioned in Sec. \ref{sec3}, the solutions we obtain are determined by $\omega$ and $q$. Once the magnetic charge $q$ is determined, a family of BPSs solutions can be obtained under the given boundary conditions. In Fig. \ref{field}, we show the distributions of the metric functions $n$ and $no^2$, as well as the Proca field functions $f$ and $g$ at a fixed magnetic charge $(q=0.8)$. From this figure, we can see that as the frequency $\omega$ decreases to 0.0002, the metric functions $n$ and $no^2$ become very small at a certain position $x_{cH}$. For an observer at infinity, the collapse process of the matter inside $x_{cH}$ appears frozen. Thus, we refer to the solution with $x_{cH}$ as the frozen Bardeen-Proca star (FBPS). And the minimum value of $n$ for FBPSs are very close to zero. Due to the similarity of the metric functions behavior at $x=x_{cH}$ to the event horizon, we refer to this surface at $x=x_{cH}$ as "critical horizon". As shown in the bottom panels of Fig. \ref{field}, near the critical horizon, the matter fields rapidly decaying. Beyond the critical horizon, the matter fields are nearly absent. In Fig. \ref{lognx}, we study the distribution of the metric function n for FBPSs ($\omega=0.0002$). We find that as $q$ increases, the critical radius $x_{cH}$ moves outward.

Fig. \ref{energy} displays the radial distribution of the Proca field energy density $\rho=-g^{0 \mu}T^{P}_{\mu 0 }$. It can be seen that when the Bardeen-Proca star becomes a frozen star at $\omega=0.0002$, the energy density will rapidly approach zero near the critical horizon. In addition, from the right panel of Fig. \ref{energy},  It can be observed that variations in the magnetic charge $q$ lead to differences in the energy density $\rho$ distribution. As $q$ increases, the maximum value of energy density $\rho_{max}$ continues to decrease. When $q$ approaches 0.883, which corresponds to the $q=3^{3/4}\sqrt{s/2}$ for $s=0.3$, $\rho_{max}$ approaches zero. If $q>0.883$, no BPSs solution exists.

\begin{figure}[!htbp]
			\begin{center}
				\includegraphics[height=.28\textheight]{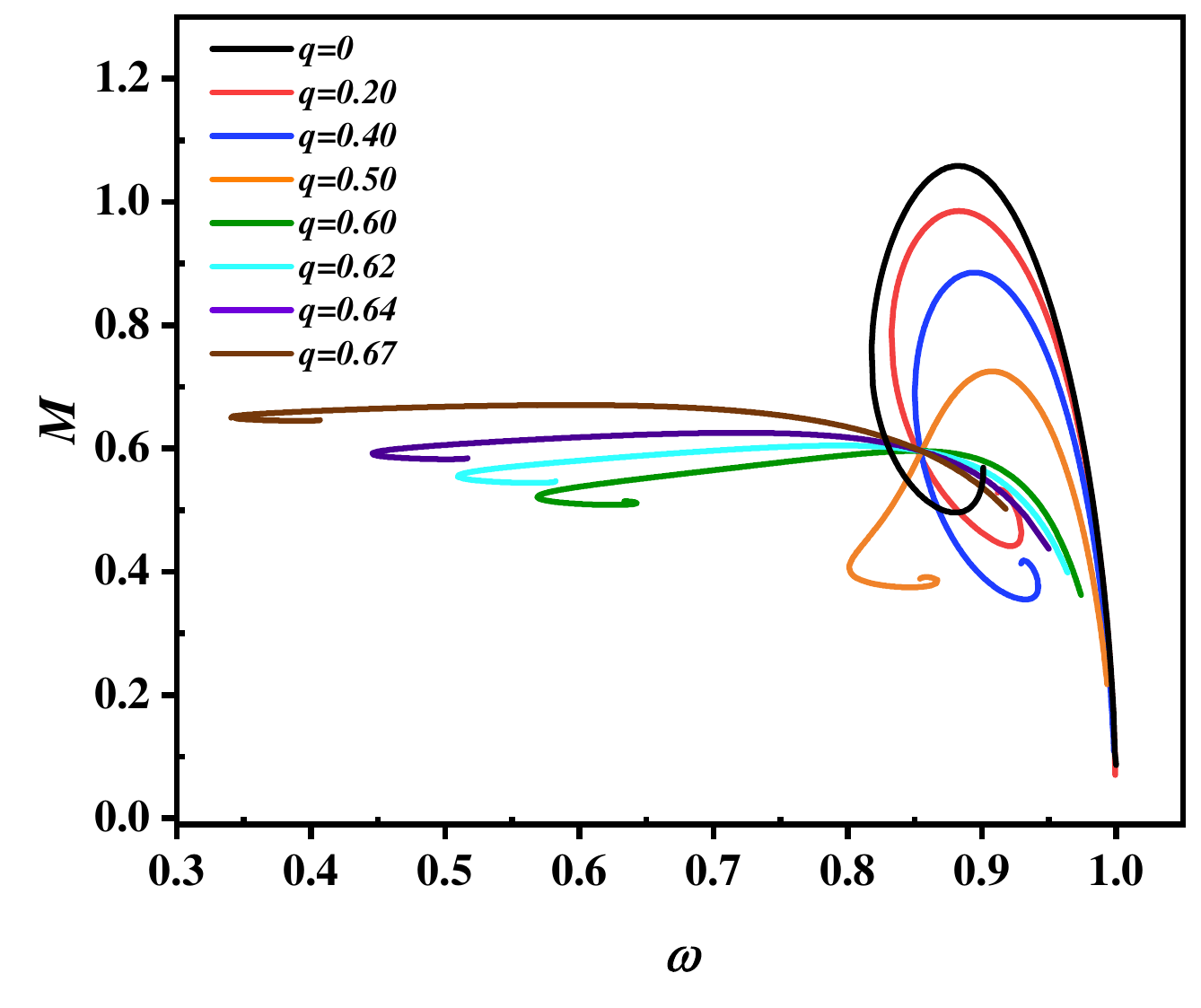}
				\includegraphics[height=.28\textheight]{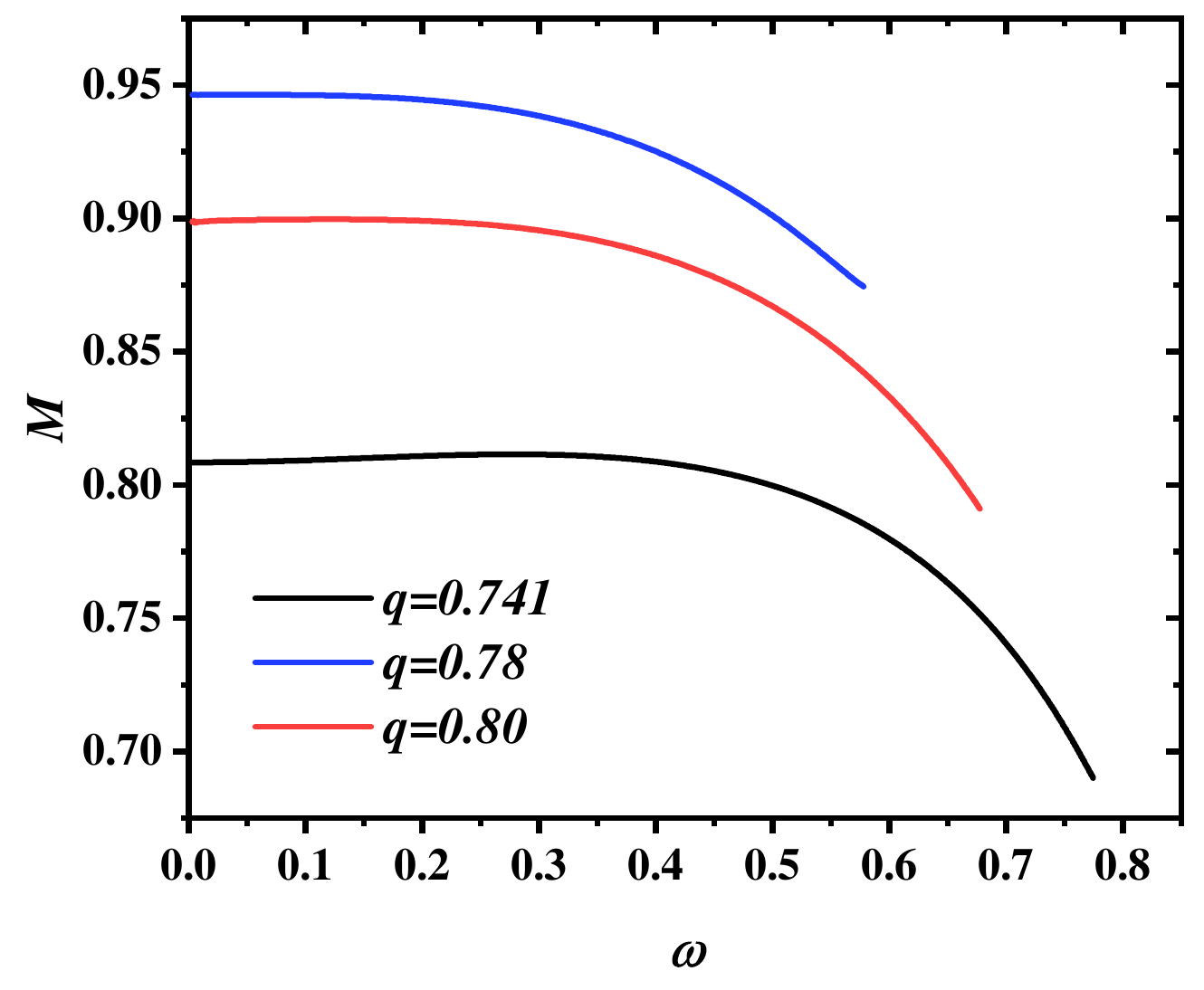}
				\includegraphics[height=.28\textheight]{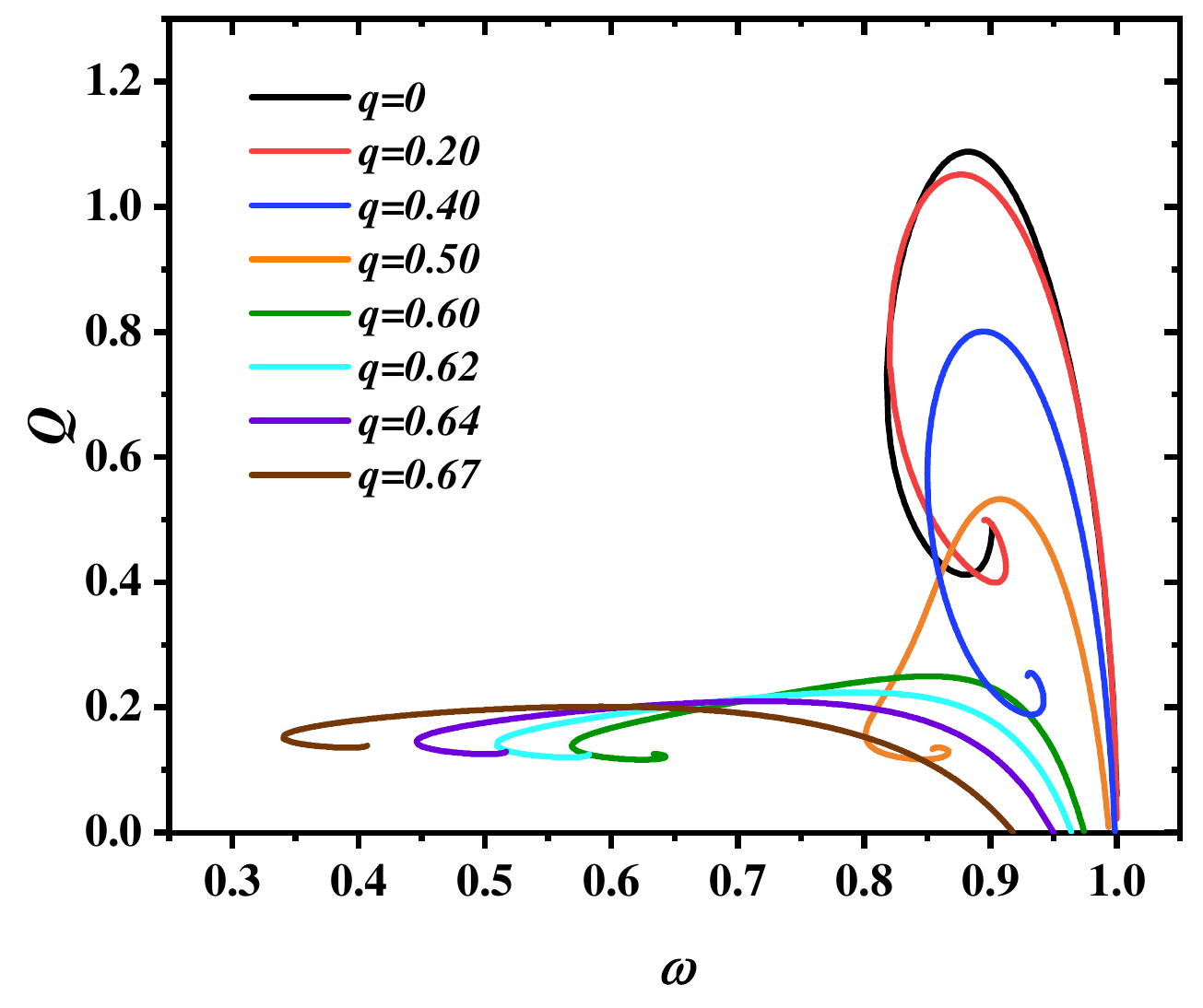}
				\includegraphics[height=.28\textheight]{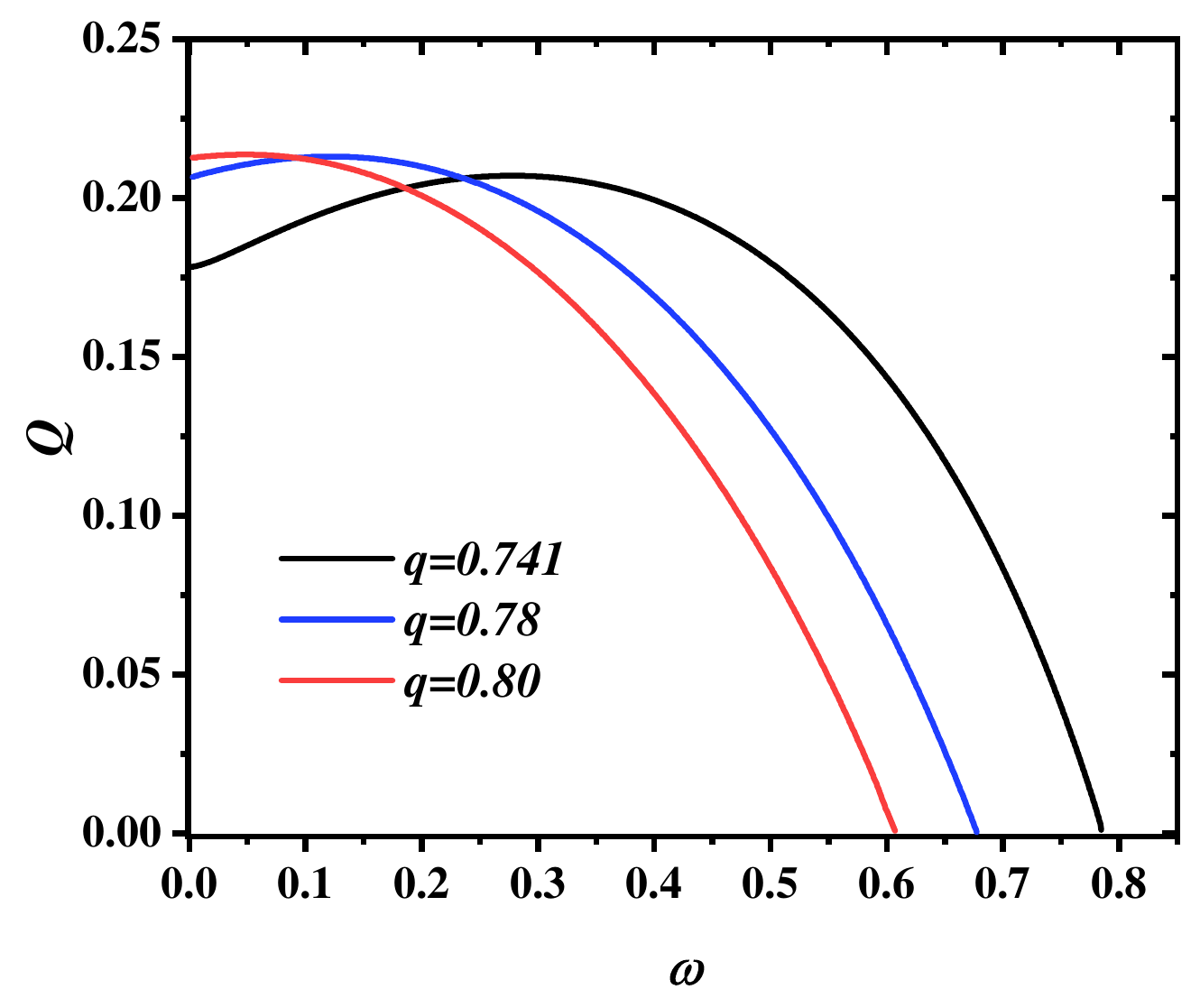}
			\end{center}
			\caption{The ADM mass $M$ and the Noether charge $Q$ of BPSs as functions of $\omega$.}
			\label{QM}
\end{figure}

In Fig. \ref{QM} we display the variation of the ADM mass $M$ and Noether charge $Q$ of BPSs as functions of frequency $\omega$. For a family of BPSs with a fixed magnetic charge $q$, each point in the parameter space $\{ \omega,M \}$ or $\{ \omega,Q \}$ represents a certain configuration. As shown in the left column of this figure, a series of spiral structures form when the magnetic charge is small. And if we continue the calculations, these curves will continue to spiral inwards, which is similar to the pure Proca stars. However, when the magnetic charge increases to 0.741, the minimum frequency $\omega$ of BPSs can decrease to very close to 0 and then terminates, with no spiral structure forming. In addition, due to the contribution of the Bardeen term, when $q>0$ the minimum mass of BPSs remains non-zero. Unlike the ADM mass $M$, the Noether charge $Q$ only represents the number of Proca particles, so its minimum value remains zero. Furthermore, both the minimum and maximum frequencies of BPSs decrease as $q$ increases.

	\begin{figure}[!htbp]
			\begin{center}
				\includegraphics[height=.28\textheight]{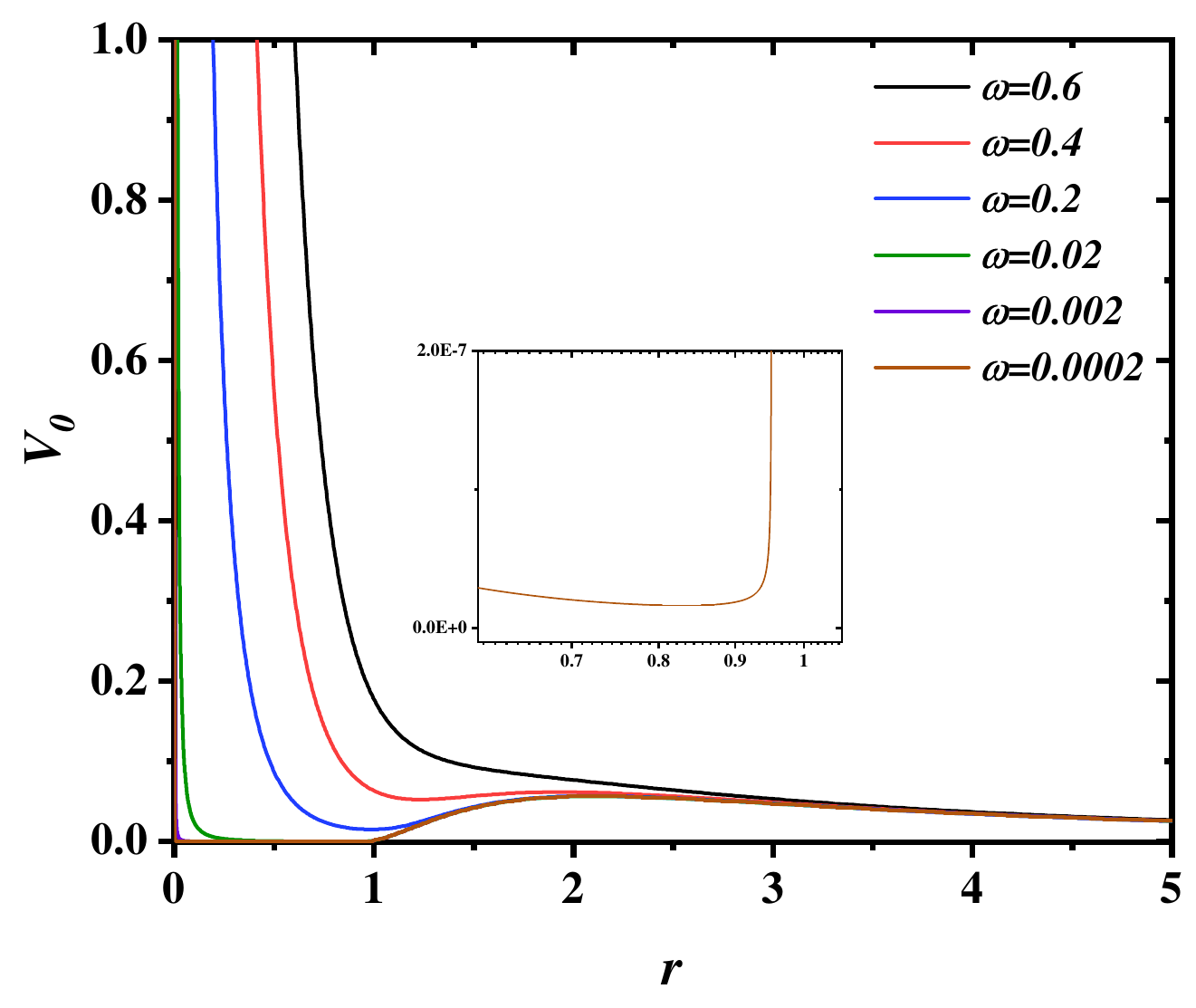}
                \includegraphics[height=.28\textheight]{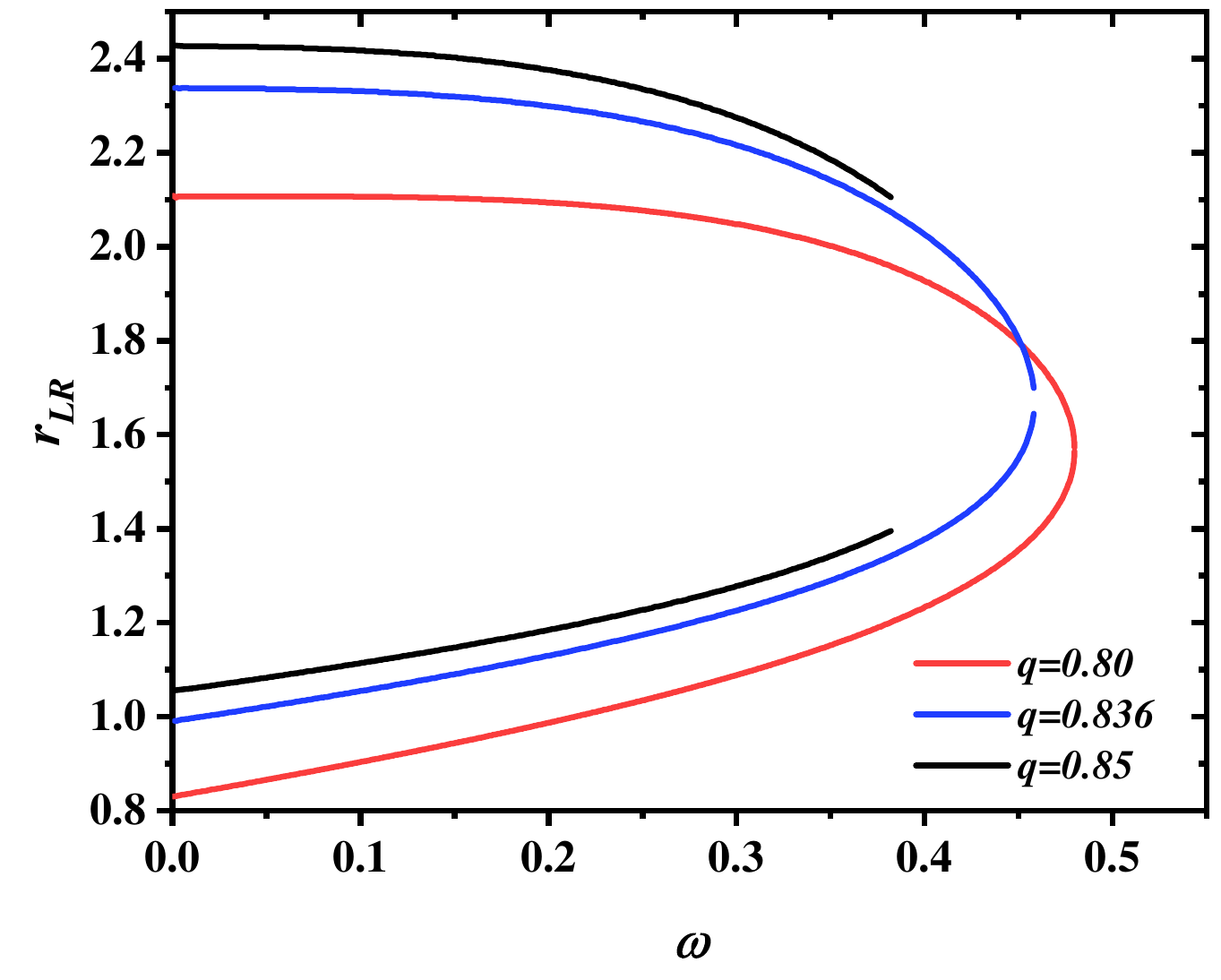}
			\end{center}
			\caption{Left panel: The distribution of the effective potential $V_{0}$ vs. the radial coordinate $r$ with $q=0.8$. Right panel: The light ring positions $r_{LR}$ of BPSs vs. the frequency $\omega$.}
			\label{DVeff}
	\end{figure}

Next, we will analyze the light rings (LRs) of BPSs. In the Sec. \ref{c}, we derived the effective potential $V_{0}$ of light ring (\ref{eq:VEFF}). Through the effective potential, we can determine the position and stability of a light ring. In the left panel of Fig. \ref{DVeff}, we displays the distribution of the effective potential $V_{0}$ along the radial coordinate $r$ for BPS with $q=0.8$. From the graph, it is shown that as the frequency $\omega$ decreases, the amount of LRs increases from zero to two. And the inner light ring is stable, while the outer light ring is unstable. Additionally, we also studied the dependence of the light ring positions $r_{LR}$ on the frequency $\omega$. The right panel of Fig. \ref{DVeff} shows the relationship between the light ring position $r_{LR}$ and the frequency $\omega$ for BPSs with $q={0.80,0.836,0.85}$. It can be seen that when the magnetic charge is fixed, as $\omega$ increases, the positions of two light rings approach each other. For $q<0.836$, two light rings finally merge into one at a critical frequency. However, for $q\geq0.836$, they no longer merge, and the minimum distance between them increases as $q$ increase.
 
\begin{figure}[!htbp]
			\begin{center}
                \includegraphics[height=.28\textheight]{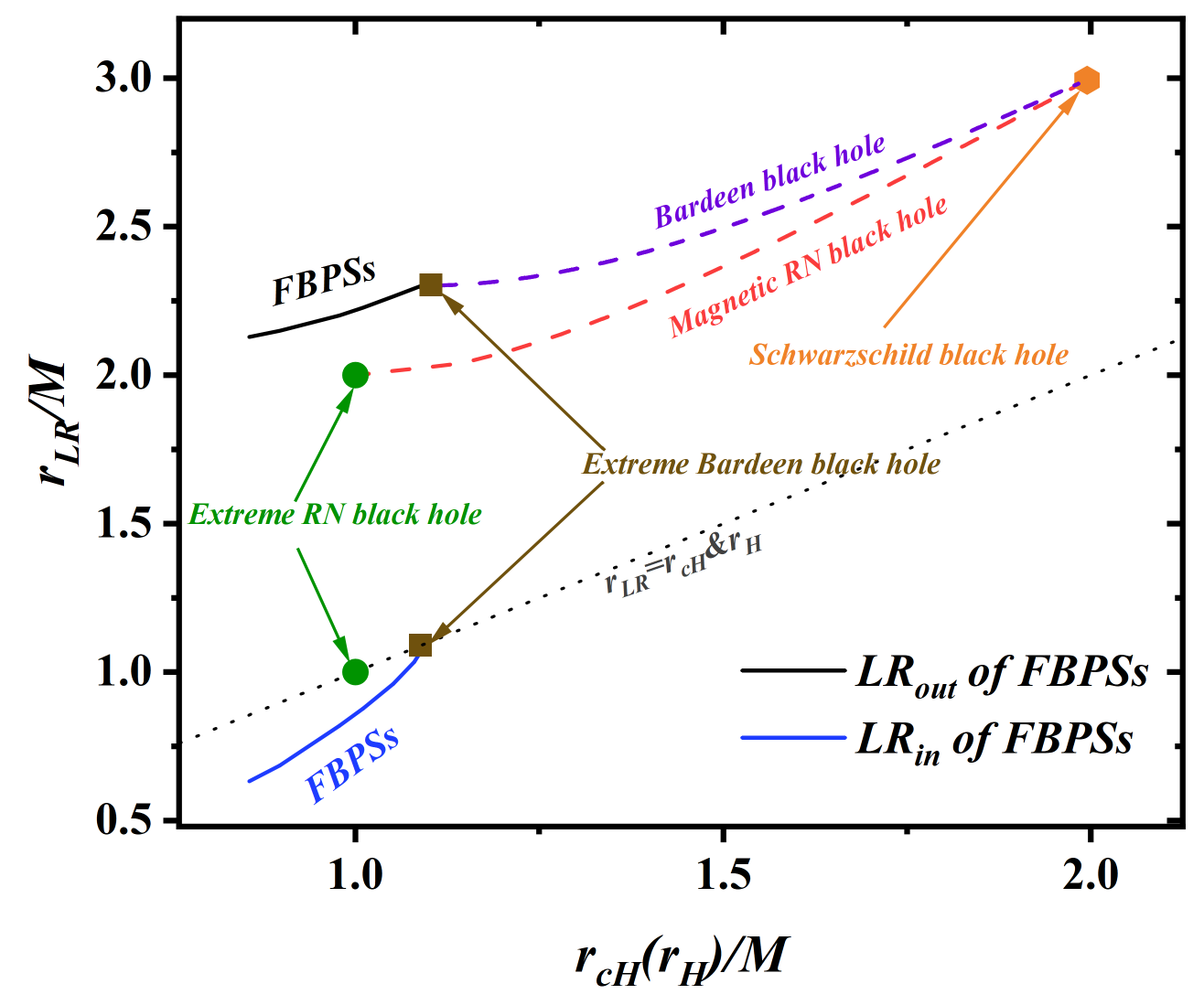}
			\end{center}
			\caption{The light ring positions $r_{LR}$ of FBPSs (black holes) vs. the position $r_{H}(r_{cH})$ of their (critical) horizons. The black dotted line indicates $r_{LR}=r_{H}(r_{cH})$.}
			\label{rLR}
	\end{figure}

Let us now focus on the properties of FBPSs as $\omega$ approaches zero. In Fig. \ref{rLR}, we present the light ring position $r_{LR}$ as a function of the critical horizon position for FBPSs. For comparison with other black hole models, we also include the magnetic Reissner–Nordström metric (RN) black hole and the Bardeen black hole in this figure. The metric of magnetic RN black hole is given by
\begin{equation}
	ds^2=-d(r)dt^2+d(r)^{-1}dr^2+r^2\left(d\theta^2+\sin^2\theta d\varphi^2\right),~d(r)=1-\frac{2M}{r}+\frac{q^2}{r^2}.
    \label{eq:RNmetric}
\end{equation}
Based on this metric (\ref{eq:RNmetric}) and effective potential $V_{0}$ (\ref{eq:VEFF}), the horizon position $r_{H}$ and light ring position $r_{LR}$ of the magnetic RN black hole can be obtained:
\begin{equation}
	r_H^{\pm}=M(1\pm\sqrt{1-\frac{q^2}{M^2}}),~r_{LR}^{\pm}=M(\frac{3}{2}\pm\frac{1}{2}\sqrt{1-\frac{q^2}{M^2}}).
\end{equation}
Using the same method, we can obtain the $r_{H}$ and $r_{LR}$ of the Bardeen black hole. We normalize the (critical) horizon and light ring positions of these three models by dividing them by the ADM mass $M$ for comparison.

It should be noted that although the effective potential of black hole models may have two extrema, due to the presence of the horizon, the light ring inside the horizon is unobservable. Only in the special cases of extreme Bardeen black holes $(q=3^{3/4}\sqrt{s/2})$ and extreme RN black holes $(q^2=M^2)$ will the inner light ring appear, coinciding with the horizon. However, unlike these two black hole models, the FBPSs have no event horizon. Therefore, light rings inside the critical horizon should be considered.

As shown in Fig. \ref{rLR}, firstly, it can be observed that the stable light rings of the FBPSs are always inside the critical horizon. As the magnetic charge $q$ approaches $3^{3/4}\sqrt{s/2}$, the FBPSs transition to the extreme Bardeen black hole, as indicated by the brown squares in the figure, while the inner light ring position $r^{-}_{LR}$ almost coincides with the critical horizon. Secondly, compared to the other two black hole models, it can be found that the smallest horizon radius of the frozen Bardeen-Proca star is smaller. And the critical horizon position $r_{cH}$ of the FBPSs is always smaller than the event horizon position $r_{H}$ of the Bardeen black hole, except for the extreme Bardeen black hole, the same relationship applies to their outer light ring positions $r^{+}_{LR}$. Furthermore, with the same $r_{H}(r_{cH})/M$, the outer light ring position $r^{+}_{LR}$ of the FBPSs is always outside that of the RN black holes.

\begin{figure}[!htbp]
	\begin{center}
		\includegraphics[height=.28\textheight]{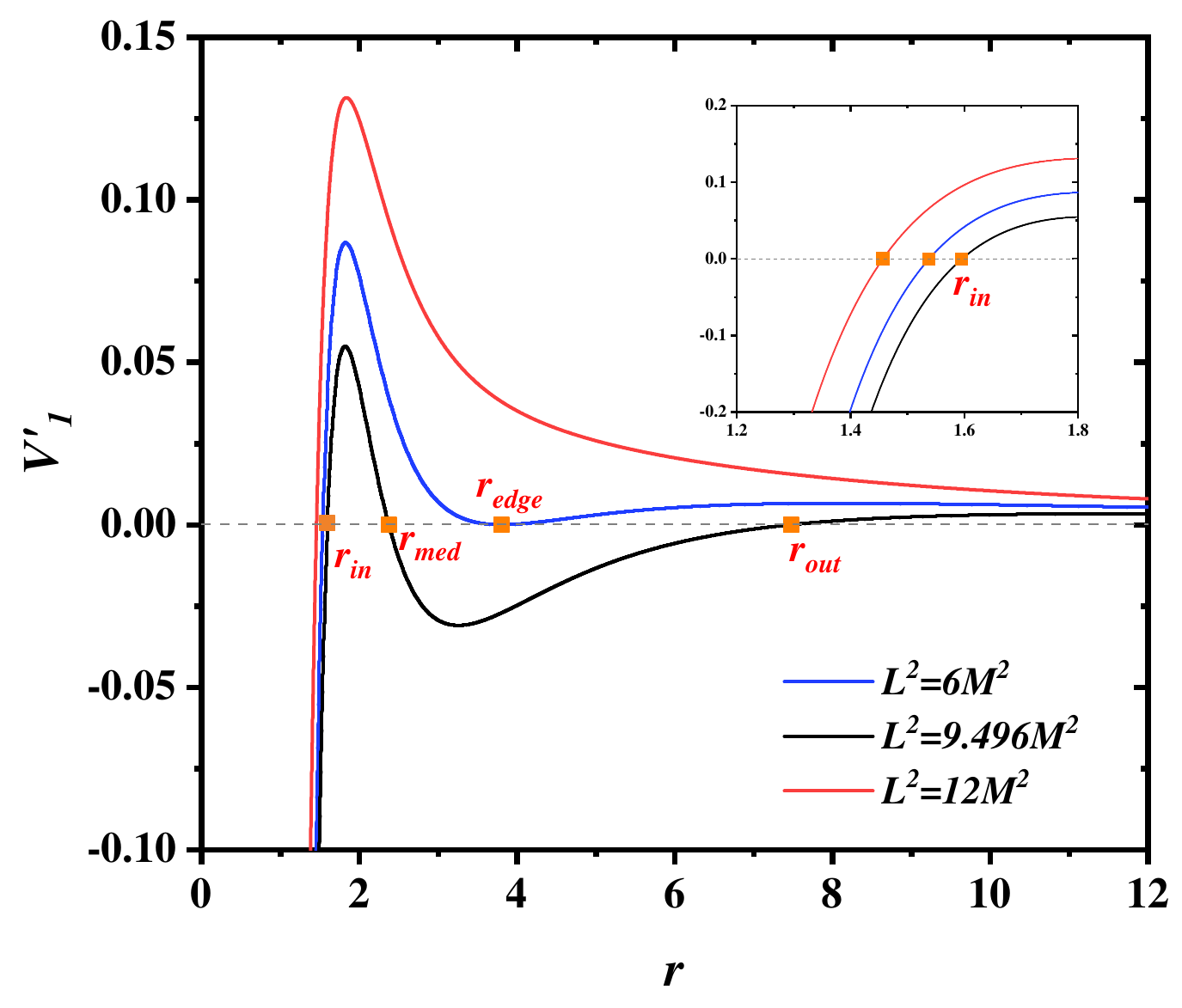}
	\end{center}
	\caption{The distribution of the derivative of the effective potential $V'_{1}$ along the radial coordinate $r$ for BPSs with $q=0.82$ and $\omega=0.531$.}
	\label{DVO82}
\end{figure}

At the end of this section, we will discuss the timelike circular orbits (TCOs) of the Bardeen-Proca stars. Unlike the null geodesics, the effective potential of timelike orbits, $V_{1}$, is related to the particle's angular momentum $L$. In Fig. \ref{DVO82}, we present the radial distribution of the derivative of the effective potential, $V'_{1}$, for Bardeen-Proca stars with $q=0.82$ and $\omega=0.531$. The different colors of the solid lines represent $L^2=\{6M^{2},9.496M^{2},12M^{2}\}$, respectively.

For particles with $L^2=6M^{2}$, the corresponding $V'_{1}$ has three zero points, resulting three timelike circular orbits, which we denote in the figure as $r_{in}$, $r_{med}$, and $r_{out}$. Among them, $r_{in}$ is unstable $(V''_{1}(r_{TL})<0)$, whereas $r_{med}$ and $r_{out}$ are stable $(V''_{1}(r_{TL})>0)$. We first focus on $r_{med}$ and $r_{out}$. As the angular momentum increases, $r_{med}$ shifts outward while $r_{out}$ moves inward. When $L^2=9.496M^{2}$, they merge at $r_{edge}$, and for $L^2>9.496M^{2}$, they disappear. Therefore, inside $r_{edge}$, there exists a region with only unstable orbits, while outside $r_{edge}$, there is a region with only stable orbits. Based on the property of the timelike circular orbits, these regions are referred to as the "Unstable TCO region" and the "Stable TCO region", respectively. As for $r_{in}$, it remains present within a finite range regardless of the changes in $L$. The space that $r_{in}$ occupies is a Stable TCO region. In fact, this stable region is adjacent to two regions where no timelike circular orbit exists \cite{Delgado:2021jxd}, which are referred to as "No TCO regions". As described in Sec. \ref{c}, the boundaries of these No TCO regions are determined by $2no^2-r(no^2)'=0$ or $(no^2)'=0$.

\begin{figure}[!htbp]
	\begin{center}
		\includegraphics[height=.28\textheight]{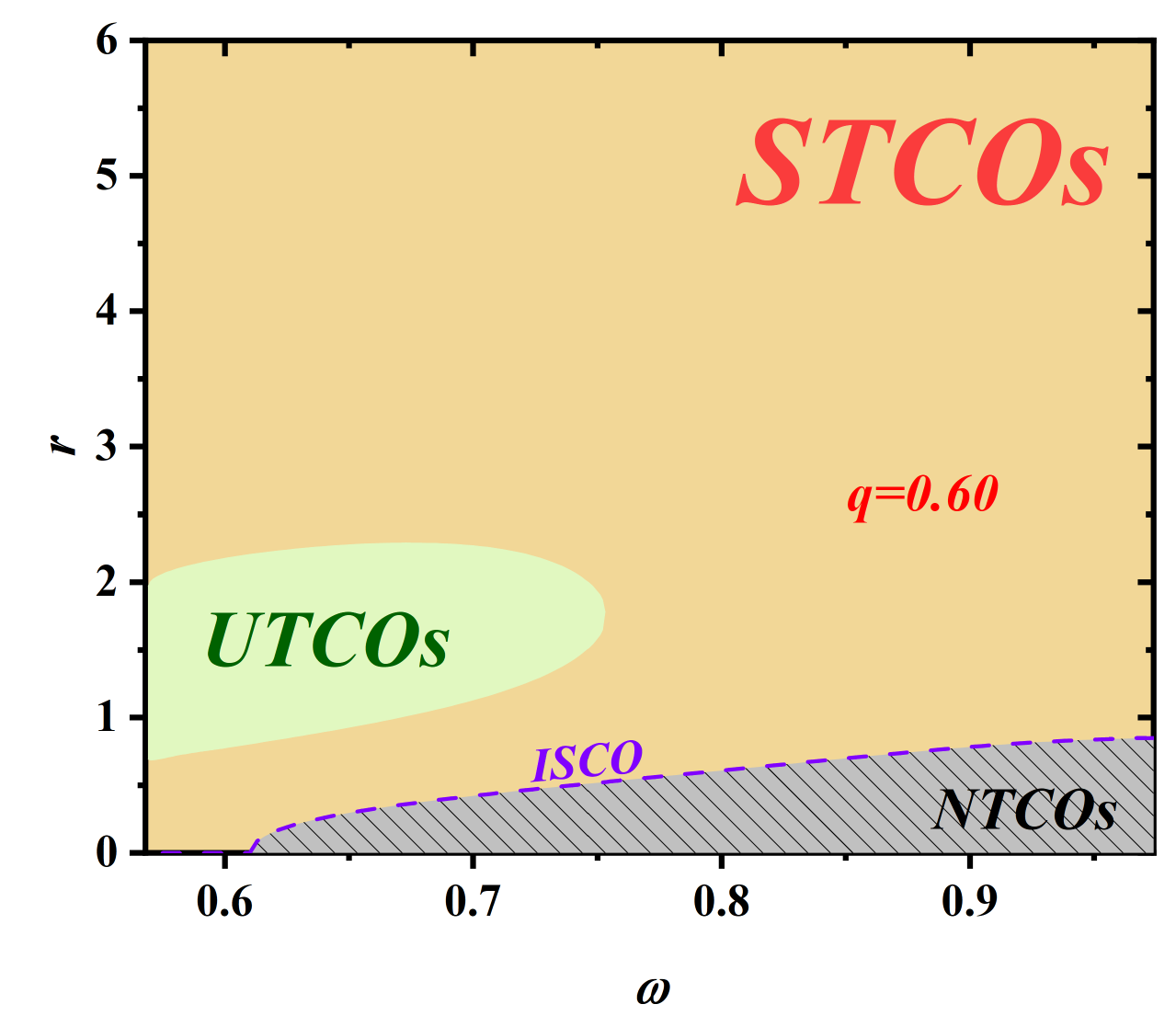}
		\includegraphics[height=.28\textheight]{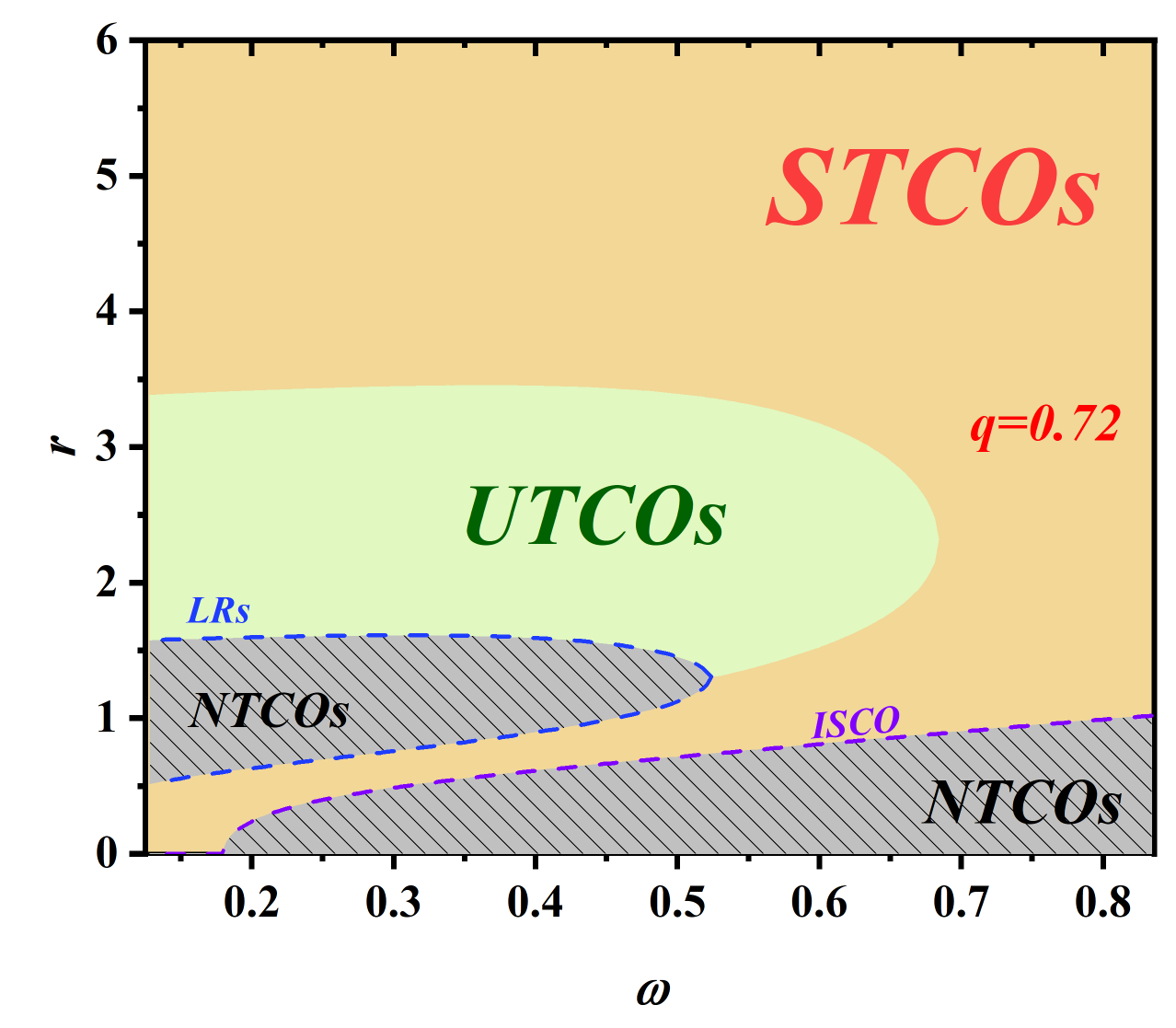}
		\includegraphics[height=.28\textheight]{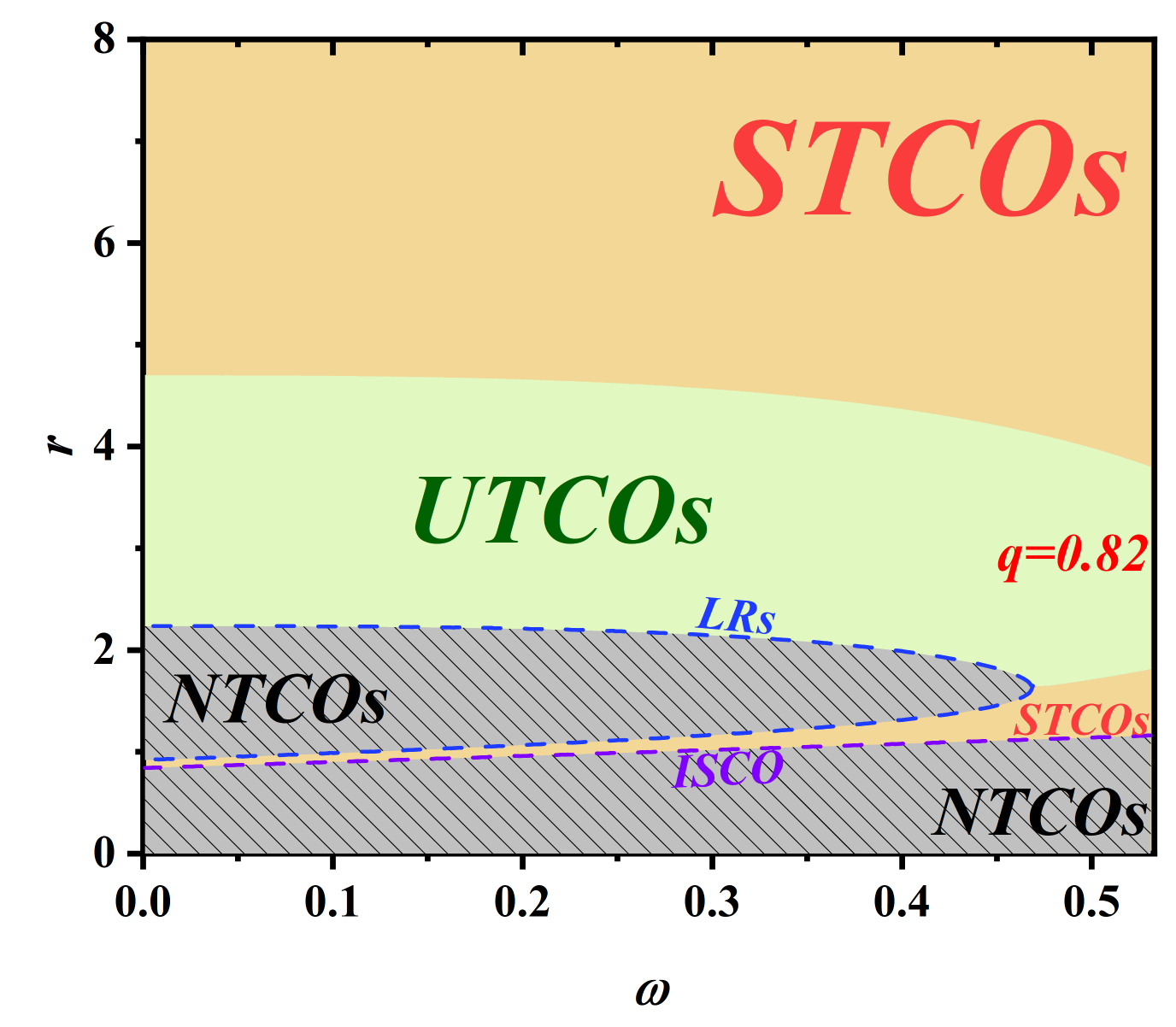}
		\includegraphics[height=.28\textheight]{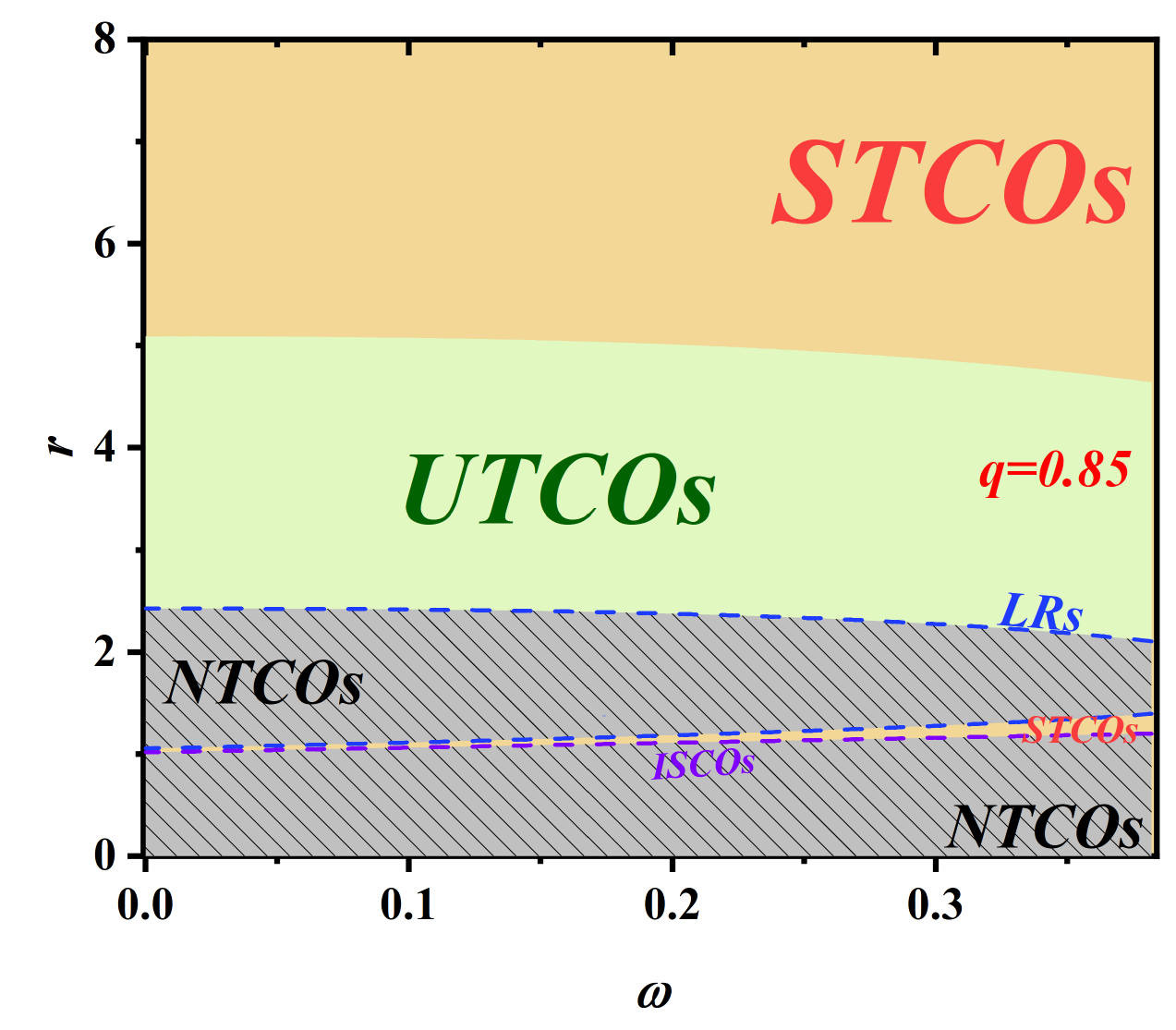}
	\end{center}
	\caption{Structure of TCOs region for BPSs with different magnetic charges $q$.}
	\label{RISCO}
\end{figure}

In Fig. \ref{RISCO}, we present the radial distribution of the TCO regions for Bardeen-Proca stars with $q=\{0.60,0.72,0.82,0.85\}$. Notably, for solution families that exhibit a spiral in the parameter space $\{\omega,M\}$ (or $\{\omega,Q\}$), only the first branch is shown in this figure. This branch includes solutions starting from the right end of these curves and following the spiral to the minimum frequency. Thus, for BPSs with a fixed $q$ in this figure, each frequency $\omega$ corresponds to a unique configuration. For a given configuration, the vertical axis represents the radial coordinates of the spacetime in which it resides. The Stable TCO region, Unstable TCO region, and No TCO region are depicted in brown, green, and gray, respectively. Furthermore, the key boundaries between these regions, namely the light rings (LRs) and the innermost stable circular orbit (ISCO), are indicated by blue and purple dashed lines, respectively.

We first focus on BPSs with smaller magnetic charge $q$. For $q=0.60$, when the frequency $\omega$ is relatively high, BPSs exhibit only a single Stable TCO region and a No TCO region. The innermost stable circular orbit is located at the boundary between these two regions, with its radial coordinate denoted by $r=r_{ISCO}$. In this scenario, the Stable TCO region forms a continuous domain, extending outward from $r=r_{ISCO}$ to spatial infinity, while the No TCO region occupies the range between $r=0$ and $r=r_{ISCO}$. As $\omega$ decreases to sufficiently low, an Unstable TCO region emerges, dividing the Stable TCO region into two separate parts. 

For BPSs with larger $q$, such as $q=0.72$ shown in the top left panel, configurations with lower frequencies will generate a pair of light rings. Since these light rings satisfy the condition $2no^2-r(no^2)'=0$, a No TCO region arises within them. This region is always located between Stable TCO region and Unstable TCO region, with its range continuously expands as the frequency decreases. It is noteworthy that, in some configurations with lower frequencies, the Stable TCO regions can extend to $r=0$, which implies the absence of the ISCO for these BPSs. For $q=\{0.82,0.85\}$, FBPs are formed as $\omega \rightarrow 0$. In the space where these FBPs exist, the structure of the TCO regions resembles that of BPSs with LRs, ISCO, and Unstable TCOs regions. Although the inner Stable TCOs region of FBPs narrows significantly as $q$ increases, it remains present and does not vanish.

\section{CONCLUSION}\label{sec5}
In this paper, we construct Bardeen-Proca Stars (BPSs), consisting of Proca field and nonlinear electromagnetic field minimally coupled with gravity. When the Proca field vanishes, this model corresponds to the Bardeen theory. And when the magnetic charge of electromagnetic field $q=0$, this model degenerates into a pure Proca star.

Unlike the Bardeen theory, when the Proca field is present, no black hole solution exists regardless of the value of $q$. If $q$ is sufficiently large, the frequency $\omega$ could appraoches zero, and the Proca field becomes concentrated within the critical horizon $r_{cH}$ and decreases sharply outside this radius. It was also found that the position of the critical horizon is related to magnetic charge. As $q$ increases, the critical horizon moves outward. 

We also investigated the circular geodesic orbits of BPSs. For light rings (LRs), we found that the light rings of BPSs always appear in pairs. They are located on either side of the critical horizon, with the inner light ring being stable and the outer light ring being unstable. As $\omega$ decreases, they move further apart. We also studied the timelike circular orbits (TCOs) of BPSs. Based on the stability and existence of timelike circular orbits, we divided the space into three regions: the Stable TCO region, the Unstable TCO region, and the No TCO region. We investigated four representative families of BPS solutions, and discussed the structure of their TCO regions. In addition, we also discussed the relationship between timelike circular orbits and light rings.

There are many directions worth further exploring in our work. For instance, it was found that using the Hayward action coupled with scalar fields can also yield frozen stars. Therefore, it is expected that the Hayward action may have the same effect on the Proca field. Additionally, the dynamical stability of these models is also a worthwhile topic for research. Finally, due to the ubiquity of rotational phenomena, we hope to extend our research to include rotating cases.

\section*{Acknowledgements}
This work is supported by the National Natural Science Foundation of China (Grant No.~12275110 and No.~12247101) and  the National Key Research and Development Program of China (Grant No. 2020YFC2201503 and Grant No. 2021YFC2203003).


\begin{thebibliography}{99}

\bibitem{EventHorizonTelescope:2019dse}
K.~Akiyama \textit{et al.} [Event Horizon Telescope],
``First M87 Event Horizon Telescope Results. I. The Shadow of the Supermassive Black Hole,''
Astrophys. J. Lett. \textbf{875}, L1 (2019)
[arXiv:1906.11238 [astro-ph.GA]].

\bibitem{EventHorizonTelescope:2019uob}
K.~Akiyama \textit{et al.} [Event Horizon Telescope],
``First M87 Event Horizon Telescope Results. II. Array and Instrumentation,''
Astrophys. J. Lett. \textbf{875}, no.1, L2 (2019)
[arXiv:1906.11239 [astro-ph.IM]].

\bibitem{EventHorizonTelescope:2019jan}
K.~Akiyama \textit{et al.} [Event Horizon Telescope],
``First M87 Event Horizon Telescope Results. III. Data Processing and Calibration,''
Astrophys. J. Lett. \textbf{875}, no.1, L3 (2019)
[arXiv:1906.11240 [astro-ph.GA]].

\bibitem{EventHorizonTelescope:2019ths}
K.~Akiyama \textit{et al.} [Event Horizon Telescope],
``First M87 Event Horizon Telescope Results. IV. Imaging the Central Supermassive Black Hole,''
Astrophys. J. Lett. \textbf{875}, no.1, L4 (2019)
[arXiv:1906.11241 [astro-ph.GA]].

\bibitem{EventHorizonTelescope:2019pgp}
K.~Akiyama \textit{et al.} [Event Horizon Telescope],
``First M87 Event Horizon Telescope Results. V. Physical Origin of the Asymmetric Ring,''
Astrophys. J. Lett. \textbf{875}, no.1, L5 (2019)
[arXiv:1906.11242 [astro-ph.GA]].

\bibitem{EventHorizonTelescope:2019ggy}
K.~Akiyama \textit{et al.} [Event Horizon Telescope],
``First M87 Event Horizon Telescope Results. VI. The Shadow and Mass of the Central Black Hole,''
Astrophys. J. Lett. \textbf{875}, no.1, L6 (2019)
[arXiv:1906.11243 [astro-ph.GA]].

\bibitem{Einstein:1916vd}
A.~Einstein,
``The foundation of the general theory of relativity.,''
Annalen Phys. \textbf{49} (1916), 769-822.

\bibitem{Schwarzschild:1916uq}
K.~Schwarzschild,
``On the gravitational field of a mass point according to Einstein's theory,''
Sitzungsber. Preuss. Akad. Wiss. Berlin (Math. Phys. ) \textbf{1916} (1916), 189-196
[arXiv:physics/9905030 [physics]].

\bibitem{Kerr:1963ud}
R.~P.~Kerr,
``Gravitational field of a spinning mass as an example of algebraically special metrics,''
Phys. Rev. Lett. \textbf{11} (1963), 237-238.

\bibitem{Newman:1965my}
E.~T.~Newman, R.~Couch, K.~Chinnapared, A.~Exton, A.~Prakash and R.~Torrence,
``Metric of a Rotating, Charged Mass,''
J. Math. Phys. \textbf{6} (1965), 918-919.

\bibitem{Penrose:1964wq}
R.~Penrose,
``Gravitational collapse and space-time singularities,''
Phys. Rev. Lett. \textbf{14}, 57-59 (1965)

\bibitem{Hawking:1970zqf}
S.~W.~Hawking and R.~Penrose,
``The Singularities of gravitational collapse and cosmology,''
Proc. Roy. Soc. Lond. A \textbf{314}, 529-548 (1970)

\bibitem{Shirokov1948}
M.~F.~Shirokov, 
``Solutions of the Schwarzschild-Nordstrom type for a point charge without singularities,''
Soviet Phys. JETP \textbf{18}, 236 (1948).

\bibitem{Duan:1954bms}
Y.~S.~Duan,
``Generalization of regular solutions of Einstein's gravity equations and Maxwell's equations for point-like charge,''
Sov. Phys. JETP \textbf{27} (1954), 756-758
[arXiv:1705.07752 [gr-qc]].

\bibitem{Sakharov:1966aja}
A.~D.~Sakharov,
``Nachal'naia stadija rasshirenija Vselennoj i vozniknovenije neodnorodnosti raspredelenija veshchestva,''
Sov. Phys. JETP \textbf{22} (1966), 241.

\bibitem{Gliner:1966}
E.~B.~Gliner,
``Algebraic Properties of the Energy-momentum Tensor and Vacuum-like States of Matter,''
Sov. Phys. JETP \textbf{22} (1966), 378.

\bibitem{Bardeen:1968}
J.~Bardeen,
``Non-singular general-relativistic gravitational collapse,''
Proc. Int. Conf. GR5, Tiflis, (1968).

\bibitem{Ayon-Beato:2000mjt}
E.~Ayon-Beato and A.~Garcia,
``The Bardeen model as a nonlinear magnetic monopole,''
Phys. Lett. B \textbf{493} (2000), 149-152
[arXiv:gr-qc/0009077 [gr-qc]].

\bibitem{Dymnikova:1992ux}
I.~Dymnikova,
``Vacuum nonsingular black hole,''
Gen. Rel. Grav. \textbf{24} (1992), 235-242.

\bibitem{Ayon-Beato:1998hmi}
E.~Ayon-Beato and A.~Garcia,
``Regular black hole in general relativity coupled to nonlinear electrodynamics,''
Phys. Rev. Lett. \textbf{80} (1998), 5056-5059
[arXiv:gr-qc/9911046 [gr-qc]].

\bibitem{Balart:2014cga}
L.~Balart and E.~C.~Vagenas,
``Regular black holes with a nonlinear electrodynamics source,''
Phys. Rev. D \textbf{90} (2014), 124045
[arXiv:1408.0306 [gr-qc]].

\bibitem{Balart:2016zrd}
L.~Balart and F.~Pe\~na,
``Regular Charged Black Holes, Quasilocal Energy and Energy Conditions,''
Int. J. Mod. Phys. D \textbf{25} (2016), 1650072
[arXiv:1603.07782 [gr-qc]].

\bibitem{Rodrigues:2018bdc}
M.~E.~Rodrigues and M.~V.~de Sousa Silva,
``Bardeen Regular Black Hole With an Electric Source,''
JCAP \textbf{06} (2018), 025
[arXiv:1802.05095 [gr-qc]].

\bibitem{deSousaSilva:2018kkt}
M.~V.~de Sousa Silva and M.~E.~Rodrigues,
``Regular black holes in $f(G)$ gravity,''
Eur. Phys. J. C \textbf{78} (2018), 638
[arXiv:1808.05861 [gr-qc]].

\bibitem{Balart:2014jia}
L.~Balart and E.~C.~Vagenas,
``Regular black hole metrics and the weak energy condition,''
Phys. Lett. B \textbf{730} (2014), 14-17
[arXiv:1401.2136 [gr-qc]].

\bibitem{Bambi:2013ufa}
C.~Bambi and L.~Modesto,
``Rotating regular black holes,''
Phys. Lett. B \textbf{721} (2013), 329-334
[arXiv:1302.6075 [gr-qc]].

\bibitem{Lan:2020fmn}
C.~Lan, Y.~G.~Miao and H.~Yang,
``Quasinormal modes and phase transitions of regular black holes,''
Nucl. Phys. B \textbf{971} (2021), 115539
[arXiv:2008.04609 [gr-qc]].

\bibitem{Bueno:2024dgm}
P.~Bueno, P.~A.~Cano and R.~A.~Hennigar,
``Regular Black Holes From Pure Gravity,''
[arXiv:2403.04827 [gr-qc]].

\bibitem{Barenboim:2024dko}
J.~Barenboim, A.~V.~Frolov and G.~Kunstatter,
``No Drama in 2D Black Hole Evaporation,''
[arXiv:2405.13373 [gr-qc]].

\bibitem{Ansoldi:2008jw}
S.~Ansoldi,
``Spherical black holes with regular center: A Review of existing models including a recent realization with Gaussian sources,''
[arXiv:0802.0330 [gr-qc]].

\bibitem{Lan:2023cvz}
C.~Lan, H.~Yang, Y.~Guo and Y.~G.~Miao,
``Regular Black Holes: A Short Topic Review,''
Int. J. Theor. Phys. \textbf{62} (2023) no.9, 202
[arXiv:2303.11696 [gr-qc]].

\bibitem{Torres:2022twv}
R.~Torres,
``Regular Rotating Black Holes: A Review,''
[arXiv:2208.12713 [gr-qc]].

\bibitem{Carballo-Rubio:2025fnc}
R.~Carballo-Rubio, F.~Di Filippo, S.~Liberati, M.~Visser, J.~Arrechea, C.~Barcel\'o, A.~Bonanno, J.~Borissova, V.~Boyanov and V.~Cardoso, \textit{et al.}
``Towards a Non-singular Paradigm of Black Hole Physics,''
[arXiv:2501.05505 [gr-qc]].
\bibitem{Li:2020ffy}
H.~B.~Li, Y.~B.~Zeng, Y.~Song and Y.~Q.~Wang,
``Self-interacting multistate boson stars,''
JHEP \textbf{04} (2021), 042
[arXiv:2006.11281 [gr-qc]].

\bibitem{Herdeiro:2020jzx}
C.~A.~R.~Herdeiro and E.~Radu,
``Asymptotically flat, spherical, self-interacting scalar, Dirac and Proca stars,''
Symmetry \textbf{12} (2020), 2032
[arXiv:2012.03595 [gr-qc]].

\bibitem{Sun:2022duv}
S.~X.~Sun, L.~Zhao and Y.~Q.~Wang,
``Chains of mini-boson stars,''
JHEP \textbf{08} (2023), 152
[arXiv:2210.09265 [gr-qc]].

\bibitem{Zhang:2021xhp}
Y.~P.~Zhang, Y.~B.~Zeng, Y.~Q.~Wang, S.~W.~Wei and Y.~X.~Liu,
``Motion of test particle in rotating boson star,''
Phys. Rev. D \textbf{105} (2022), 044021
[arXiv:2107.04848 [gr-qc]].

\bibitem{Kleihaus:2009kr}
B.~Kleihaus, J.~Kunz, C.~Lammerzahl and M.~List,
``Charged Boson Stars and Black Holes,''
Phys. Lett. B \textbf{675} (2009), 102-115
[arXiv:0902.4799 [gr-qc]].

\bibitem{Guzman:2009zz}
F.~S.~Guzman and J.~M.~Rueda-Becerril,
``Spherical boson stars as black hole mimickers,''
Phys. Rev. D \textbf{80} (2009), 084023
[arXiv:1009.1250 [astro-ph.HE]].

\bibitem{Cunha:2015yba}
P.~V.~P.~Cunha, C.~A.~R.~Herdeiro, E.~Radu and H.~F.~Runarsson,
``Shadows of Kerr black holes with scalar hair,''
Phys. Rev. Lett. \textbf{115} (2015), 211102
[arXiv:1509.00021 [gr-qc]].

\bibitem{Cardoso:2019rvt}
V.~Cardoso and P.~Pani,
``Testing the nature of dark compact objects: a status report,''
Living Rev. Rel. \textbf{22} (2019), 4
[arXiv:1904.05363 [gr-qc]].
\bibitem{Kaup:1968zz}
D.~J.~Kaup,
``Klein-Gordon Geon,''
Phys. Rev. \textbf{172} (1968), 1331-1342.

\bibitem{Ruffini:1969qy}
R.~Ruffini and S.~Bonazzola,
``Systems of selfgravitating particles in general relativity and the concept of an equation of state,''
Phys. Rev. \textbf{187} (1969), 1767-1783.
\bibitem{Brito:2015pxa}
R.~Brito, V.~Cardoso, C.~A.~R.~Herdeiro and E.~Radu,
``Proca stars: Gravitating Bose\textendash{}Einstein condensates of massive spin 1 particles,''
Phys. Lett. B \textbf{752} (2016), 291-295
[arXiv:1508.05395 [gr-qc]].

\bibitem{SalazarLandea:2016bys}
I.~Salazar Landea and F.~Garc\'\i{}a,
``Charged Proca Stars,''
Phys. Rev. D \textbf{94} (2016), 104006
[arXiv:1608.00011 [hep-th]].

\bibitem{Su:2023zhh}
X.~Su, C.~H.~Hao, J.~R.~Ren and Y.~Q.~Wang,
``Proca stars in wormhole spacetime,''
JCAP \textbf{09} (2024), 010
[arXiv:2311.17557 [gr-qc]].

\bibitem{Zhang:2023rwc}
R.~Zhang, L.~X.~Huang and Y.~Q.~Wang,
``Rotating multistate Proca stars,''
[arXiv:2312.15755 [gr-qc]].

\bibitem{Finster:1998ws}
F.~Finster, J.~Smoller and S.~T.~Yau,
``Particle - like solutions of the Einstein-Dirac equations,''
Phys. Rev. D \textbf{59} (1999), 104020
[arXiv:gr-qc/9801079 [gr-qc]].

\bibitem{Finster:1998ux}
F.~Finster, J.~Smoller and S.~T.~Yau,
``Particle - like solutions of the Einstein-Dirac-Maxwell equations,''
Phys. Lett. A \textbf{259} (1999), 431-436
[arXiv:gr-qc/9802012 [gr-qc]].

\bibitem{Dzhunushaliev:2018jhj}
V.~Dzhunushaliev and V.~Folomeev,
``Dirac stars supported by nonlinear spinor fields,''
Phys. Rev. D \textbf{99} (2019), 084030
[arXiv:1811.07500 [gr-qc]].

\bibitem{Huang:2023glq}
L.~X.~Huang, S.~X.~Sun, R.~Zhang, C.~Liang and Y.~Q.~Wang,
``Excited Dirac stars with higher azimuthal harmonic~index,''
JCAP \textbf{04} (2024), 085
[arXiv:2309.16497 [gr-qc]].

\bibitem{Hao:2023igi}
C.~H.~Hao, S.~X.~Sun, L.~X.~Huang, R.~Zhang, X.~Su and Y.~Q.~Wang,
``Dirac stars in wormhole spacetime,''
JCAP \textbf{04} (2024), 057
[arXiv:2309.16379 [gr-qc]].



\bibitem{Wang:2023tdz}
X.~E.~Wang,
``From Bardeen-boson stars to black holes without event horizon,''
[arXiv:2305.19057 [gr-qc]].

\bibitem{Yue:2023sep}
Y.~Yue and Y.~Q.~Wang,
``Frozen Hayward-boson stars,''
[arXiv:2312.07224 [gr-qc]].

\bibitem{Chen:2024bfj}
J.~R.~Chen and Y.~Q.~Wang,
``Hayward spacetime with axion scalar field,''
[arXiv:2407.17278 [hep-th]].

\bibitem{Huang:2023fnt}
L.~X.~Huang, S.~X.~Sun and Y.~Q.~Wang,
``Frozen Bardeen-Dirac stars and light ball,''
[arXiv:2312.07400 [gr-qc]].


\bibitem{Huang:2024rbg}
L.~X.~Huang, S.~X.~Sun and Y.~Q.~Wang,
``Bardeen spacetime with charged scalar field,''
[arXiv:2407.11355 [gr-qc]].

\bibitem{Zhang:2024ljd}
X.~Y.~Zhang, L.~Zhao and Y.~Q.~Wang,
``Bardeen-Dirac Stars in AdS Spacetime,''
[arXiv:2409.14402 [gr-qc]].

\bibitem{Sun:2024mke}
S.~X.~Sun, L.~X.~Huang, Z.~H.~Zhao and Y.~Q.~Wang,
``Bardeen Spacetime with Charged Dirac Field,''
[arXiv:2411.12969 [gr-qc]].

\bibitem{Zhao:2025hdg}
Z.~H.~Zhao, Y.~Gu, S.~Liu, L.~X.~Huang and Y.~Q.~Wang,
``Non-topological soliton Bardeen boson stars and its frozen state,''
[arXiv:2502.14153 [gr-qc]].

\bibitem{Ma:2024olw}
T.~X.~Ma and Y.~Q.~Wang,
``Frozen boson stars in an infinite tower of higher-derivative gravity,''
[arXiv:2406.08813 [gr-qc]].

\bibitem{Wang:2024ehd}
Y.~Q.~Wang,
``Frozen gravitational stars in Einsteinian cubic gravity,''
[arXiv:2410.04575 [gr-qc]].

\bibitem{Sanchis-Gual:2017bhw}
N.~Sanchis-Gual, C.~Herdeiro, E.~Radu, J.~C.~Degollado and J.~A.~Font,
``Numerical evolutions of spherical Proca stars,''
Phys. Rev. D \textbf{95} (2017) no.10, 104028
[arXiv:1702.04532 [gr-qc]].

\bibitem{Sanchis-Gual:2019ljs}
N.~Sanchis-Gual, F.~Di Giovanni, M.~Zilh\~ao, C.~Herdeiro, P.~Cerd\'a-Dur\'an, J.~A.~Font and E.~Radu,
``Nonlinear Dynamics of Spinning Bosonic Stars: Formation and Stability,''
Phys. Rev. Lett. \textbf{123}, no.22, 221101 (2019)
[arXiv:1907.12565 [gr-qc]].

\bibitem{DiGiovanni:2020ror}
F.~Di Giovanni, N.~Sanchis-Gual, P.~Cerd\'a-Dur\'an, M.~Zilh\~ao, C.~Herdeiro, J.~A.~Font and E.~Radu,
``Dynamical bar-mode instability in spinning bosonic stars,''
Phys. Rev. D \textbf{102}, no.12, 124009 (2020)
[arXiv:2010.05845 [gr-qc]].

\bibitem{CalderonBustillo:2020fyi}
J.~Calder\'on Bustillo, N.~Sanchis-Gual, A.~Torres-Forn\'e, J.~A.~Font, A.~Vajpeyi, R.~Smith, C.~Herdeiro, E.~Radu and S.~H.~W.~Leong,
``GW190521 as a Merger of Proca Stars: A Potential New Vector Boson of $8.7\times 10^{-13}$  eV,''
Phys. Rev. Lett. \textbf{126}, no.8, 081101 (2021)
[arXiv:2009.05376 [gr-qc]].

\bibitem{Rosa:2022toh}
J.~L.~Rosa, P.~Garcia, F.~H.~Vincent and V.~Cardoso,
``Observational signatures of hot spots orbiting horizonless objects,''
Phys. Rev. D \textbf{106} (2022) no.4, 044031
[arXiv:2205.11541 [gr-qc]].

\bibitem{Rosa:2022tfv}
J.~L.~Rosa and D.~Rubiera-Garcia,
``Shadows of boson and Proca stars with thin accretion disks,''
Phys. Rev. D \textbf{106} (2022) no.8, 084004
[arXiv:2204.12949 [gr-qc]].

\bibitem{Herdeiro:2021lwl}
C.~A.~R.~Herdeiro, A.~M.~Pombo, E.~Radu, P.~V.~P.~Cunha and N.~Sanchis-Gual,
``The imitation game: Proca stars that can mimic the Schwarzschild shadow,''
JCAP \textbf{04} (2021), 051
[arXiv:2102.01703 [gr-qc]].

\bibitem{Sengo:2024pwk}
I.~Sengo, P.~V.~P.~Cunha, C.~A.~R.~Herdeiro and E.~Radu,
``The imitation game reloaded: effective shadows of dynamically robust spinning Proca stars,''
JCAP \textbf{05} (2024), 054
[arXiv:2402.14919 [gr-qc]].

\bibitem{Cardoso:2008bp}
V.~Cardoso, A.~S.~Miranda, E.~Berti, H.~Witek and V.~T.~Zanchin,
``Geodesic stability, Lyapunov exponents and quasinormal modes,''
Phys. Rev. D \textbf{79} (2009) no.6, 064016
[arXiv:0812.1806 [hep-th]].

\bibitem{Delgado:2021jxd}
J.~F.~M.~Delgado, C.~A.~R.~Herdeiro and E.~Radu,
``Equatorial timelike circular orbits around generic ultracompact objects,''
Phys. Rev. D \textbf{105} (2022) no.6, 064026
[arXiv:2107.03404 [gr-qc]].

\end{thebibliography}
\end{document}